# Information: to Harvest, to Have and to Hold *

## Marin van Heel [1,2,3] ** and Michael Schatz [4]

**corresponding author

[1] Brazilian Nanotechnology National Laboratory (LNNano), Brazilian Centre for Research in Energy and Materials (CNPEM), 13083-970 Campinas, São Paulo, Brazil.
[2] Emeritus Professor, Leiden University, Biology Department (NeCEN), The Netherlands.
[3] Emeritus Professor of Structural Biology, Imperial College London, London, UK.
[4] Image Science Software GmbH, Berlin, Germany.

## Abstract

Signal-to-Noise Ratios (*SNR*s) and the Shannon-Hartley channel capacity are metrics that help define the ***loss of known information*** while transferring data through a noisy channel. These metrics cannot be used for quantifying the opposite process: the ***harvesting* of *new information***. Correlation functions and correlation coefficients do play an important role in collecting new information from noisy sources. However, Bershad and Rockmore [Bershad & Rockmore, 1974] based their correlation-to-SNR formulas on *a priori* assumptions in Real-space and in Fourier-space, which cannot be fulfilled simultaneously. Their formulations were subsequently copied literally to the practical science of electron microscopy, where those *a priori* assumptions now distort most quality metrics in Cryogenic Electron Microscopy (cryo-EM). Cryo-EM became a great success in recent years [Wiley Award 2017; Nobel prize for Chemistry 2017] and became the method of choice for revealing structures of biological complexes like ribosomes, viruses, or corona-virus spikes, vitally important during the current COVID-19 pandemic. Those early misconceptions, however, now interfere with the objective comparison of independently obtained results, especially where it concerns local details. We found that the roots of these problems significantly pre-date those 1970s publications and were already inherent in the original *SNR* definitions, introduced more than a century ago. We here propose novel metrics to assess the amount of information harvested in an experiment, information which is measured in ***bits***. These new metrics assess the total amount of information collected on an object, as well as the information density distribution within that object. The new metrics can be applied everywhere where data is collected, processed, compressed, or compared. As an example, we compare the structures of two recently published SARS-CoV-2 spike proteins. We also introduce new metrics for transducer-quality assessment in many sciences including: cryo-EM, biomedical imaging, microscopy, signal processing, photography, tomography, etc.

*This paper is dedicated to the memory of **Professor James Barber (1940-2020)**

# I) **<u>Introduction</u>**

We here discuss the collection of new information on objects present in noisy data. Such information harvesting has not been properly integrated into ***information theory***. The Signal-to-Noise Ratio (***SNR***) and Shannon's information capacity concepts are focused on *not losing* the information contained in a message while transferring that message through a ***channel***. The channel can be a physical telephone line or an abstract operation like the storing and retrieving of messages from a computer hard-drive. The very concept of a *mathematical theory of communication* [Shannon 1948], implies one knows all information deterministically at point **A** that one wants to send through the channel to point **B**. If there is no signal entering the channel at **A**, there is also no way for an observer at **B** to know from the arriving noise that it does *not* contain some hidden signal and that thus the ***SNR*** is zero. The observer needs to have the *a priori* knowledge that no signal was sent from **A**.

Collecting new information from a noisy source is an entirely different matter. The ***signal*** enters a transducer in the form of ***waves*** which are detected as ***intensities***, and integrated over a certain time in order to form a ***measurement*** (a ***measurement vector***). The ***integration time*** $\tau$ is an important concept: when using a very short integration time, the measurement vectors will be very noisy and the cross-correlation coefficient (***CCC***) between different measurement vectors will be small and noisy. What we want to emphasise here is that the ***CCC*** is not an intrinsic property of the source, but rather a property of the measurement. ***CCC***-based approaches can be used to harvest new information which is not possible with the ***SNR***-oriented approaches. One can use cross-correlations to search for similarities between two independent Real-space measurements from the same source. Correlations can also be used to find multiple copies of signals in noisy measurements. Once identified, one can compare such signals (or averaged signals) by Real-space ***CCCs***.

The ***CCC***s in Real-space are, however, mostly only of limited use in 2D or 3D data analysis, due to the typical overwhelming presence of low-frequency components in normal images [Van Heel 1992]. It makes more sense to study ***CCC***s in Fourier-space as function of spatial frequency. Fourier-space metrics like the Fourier Ring Correlation (***FRC***) [Van Heel 1982; Saxton 1982] and the Fourier Shell Correlation (***FSC***) [Harauz 1986] are now used routinely to assess the similarity between two images or two 3D volumes. Originally used primarily in electron microscopy, they have now proliferated into most fields of scientific imaging (see: [Baksh 2020; Donnelly 2020; Loetgering 2020]). The ***CCC***s in Fourier-space are measured over rings in 2D Fourier-space (***FRC***), or shells in 3D Fourier-space (***FSC***). The ***FRC***/***FSC*** cross-correlation coefficients, are neither ***SNR***s nor ***information*** in the sense of Shannon. Their values increase when more data is accumulated, but the question remains: how to integrate these metrics into the world of ***SNR***s, and Shannon's information concepts?

To get to the heart of the matter, we will first cover some underlying fundamental signal- and image-processing concepts, starting with the more historical ones, to then focus on more recent methodological insights and developments. We will need this basis to evaluate the current state of affairs, and to then be able to build upon sound foundations.

## II) Persistent early-days signal-processing flaws

The Signal-to-Noise Ratio (**SNR**) of a dataset, defined as the ratio of the signal power over noise power: $\mathbf{SNR = s^2/n^2}$, is not a directly measurable entity, but **SN**Rs are nevertheless generally accepted as important metrics for estimating the information content of a measurement. In fact, the **SNR** is either known *a priori*, say, in a model experiment, or can otherwise at best be *estimated*, since the signal cannot be measured separately from the noise in an incoming measurement. In contrast, the normalised Cross-Correlation Coefficient (**CCC**), also known as the Pearson correlation coefficient, between two measurements, can always be determined and returns a normalised value between **-1** and **+1**. In the absence of noise, the values **+1** or **-1** will result for two fully identical measurements, and for two fully anti-correlating measurements, respectively. In the absence of a signal, the **CCC** between two noisy measurements will oscillate around the **0** mark. In the case of the **SNR**, we need to know *a priori* that no signal is present in the measurement in order to conclude that the measurement represents pure random noise, and the associated **SNR** is thus zero!

The **SNR** being so difficult to assess, Bershad and Rockmore [1974] suggested to estimate the **SNR** indirectly from the always accessible **CCC** between subsequent measurements. These authors considered the case that the same signal $\boldsymbol{s(t)}$ was measured twice, each deteriorated by different realisations of additive random noise, $\boldsymbol{n_1(t)}$ and $\boldsymbol{n_2(t)}$, to yield two measurement vectors: $\boldsymbol{x(t)}$, and $\boldsymbol{y(t)}$, respectively (**Fig 1A**). The measurements $\boldsymbol{x(t)}$, and $\boldsymbol{y(t)}$ were also assumed to be band-limited with a maximum bandwidth of $\boldsymbol{B}$ (**Fig 1A**). A limited bandwidth implies that the sampling for these measurements must be performed at a sampling frequency higher than $2\boldsymbol{B}$, in adherence to the Shannon-Nyquist sampling rules for 1D data. These sampling rules also imply that the power-spectrum of both $\boldsymbol{x(t)}$ and $\boldsymbol{y(t)}$, (and their signal and additive noise vector components), cannot be white but must have gradually dropped to zero *prior* to reaching the Nyquist frequency.

At the same time, Bershad and Rockmore assume that the expectation value of the cross-correlations between all different elements, including those between *neighbouring* elements, to be zero (**Fig 1A**). In their own words: $\langle x_i x_j \rangle = (\boldsymbol{P_s + P_n})\boldsymbol{\delta_{ij}}$ (their formula (2)). When the signal and noise vectors are band-limited in Fourier-space, that implies that *neighbouring* elements in Real-space are correlated; framing that in their notation: $\langle x_i x_{i+1} \rangle \neq \boldsymbol{0}$, and $\langle y_i y_{i+1} \rangle \neq \boldsymbol{0}$. All cross-correlation terms between neighbours (in 1D) therefore can also not be zero: $\langle x_i y_{i+1} \rangle \neq \boldsymbol{0}$. The *a priori* assumptions made by Bershad and Rockmore, in Fourier-space and in Real-space are thus mutually contradictory! There simply cannot be a sufficiently-sampled Real-space measurement (in the sense of the Shannon-Nyquist sampling theorem), where neighbouring sampling points are uncorrelated. There thus cannot be any real-life application of this mathematical/statistical theory. (Side remark: the smaller the number of samples $\boldsymbol{N}$ in a measurement, the larger the relative number of close-to-Nyquist elements, and the more serious the violation of these *a priori* assumptions become.) Note that, since the measurements are assumed to contain a fixed signal and additive random noise, $\boldsymbol{x(t) = s(t) + n_1(t)}$, the inner-products between measurement vectors generate signal versus noise cross terms that must be assessed individually.

Indicative of the fundamental problems with the Bershad and Rockmore *CCC*-to-*SNR* relation is that their final formula is fatally flawed (***SNR=CCC*/(1-*CCC*)**; **Fig 1B**). Whereas the *SNR* is – per definition – positive, the *CCC* can assume values from: **-1 ≤ *CCC* ≤ +1**. In the absence of a signal, for example, the *CCC* oscillates around zero. Any negative value for the *CCC* in this formula, directly translates to a negative value for the resulting *SNR* value, incompatible with its very definition. This *CCC*-to-*SNR* relation therefore could, at best, only give reasonable *SNR* approximations in the case where the *CCC* is very close to positive unity, that is, for data with a very high *SNR* value. In information harvesting, however, we are primarily interested in the case where a small signal emerges from a noisy background. It is in that limit that the *CCC*-to-*SNR* relation fails completely.

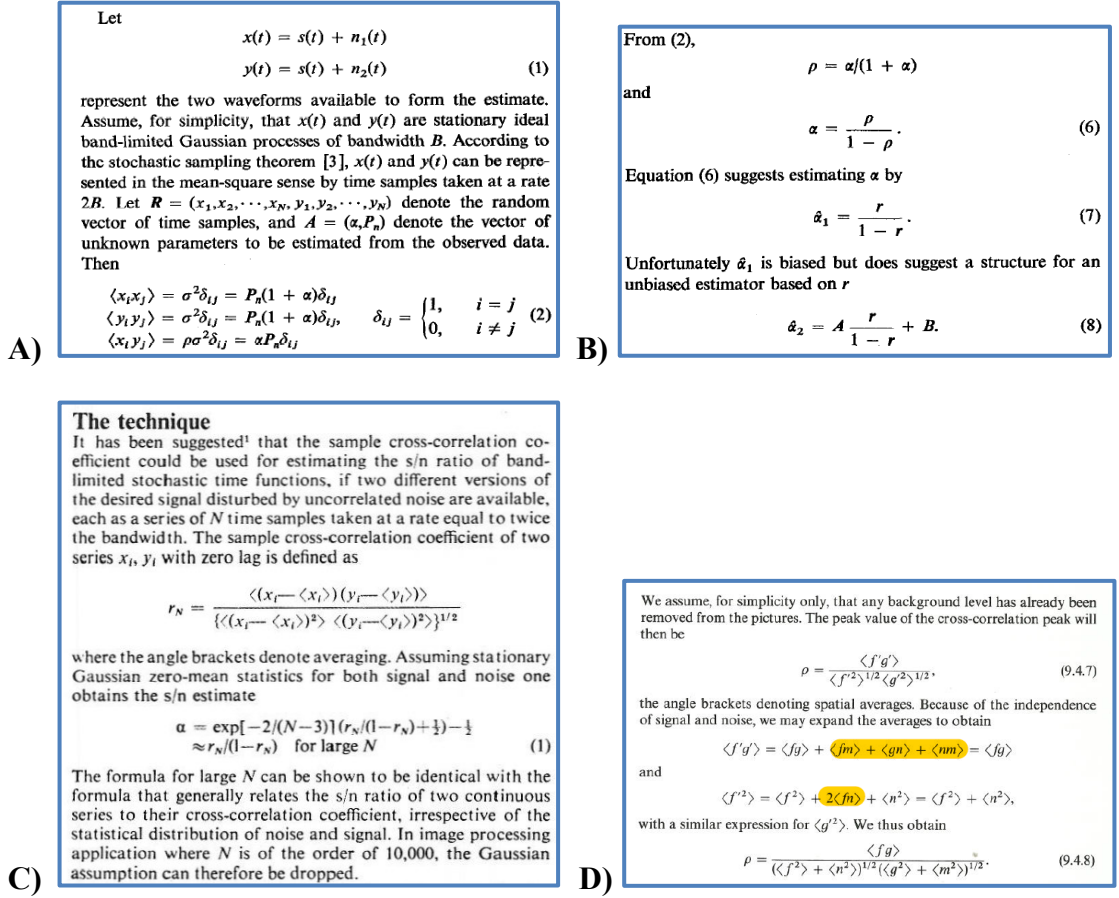

**Figure 1**: **Excerpts from the original *CCC*-to-*SNR* papers.**
We reproduced these exact excerpts to show the literal form in which *a priori* assumptions are made in these papers: (**A-B**) [Bershad & Rockmore 1974]; and in the follow-up papers: (**C**) [Frank & Al-Ali 1975] and (**D**) [Saxton 1978]. In the latter paper the explicitly deleted cross-terms are marked in yellow.

This *CCC*-to-*SNR* formula was then copied literally to microscopy in [Frank & Al-Ali, 1975], thus violating the (impossibly) strict boundary conditions of [Bershad & Rockmore 1974]. No justifications were given for this extended applicability other than: "*In image* *processing application where N is of the order of 10,000, the Gaussian assumption can therefore be dropped*" (**Fig 1C**). Saxton [Saxton 1978] then provided a comprehensive derivation of the ***CCC***-to-***SNR*** formula, pinpointing precisely the cross-terms that were assumed to be zero ("*because of the independence of signal and noise*") and which were taken out of the equations (**Fig 1D**). In all three papers, the same fundamental mistake was made: concluding from a ***zero-expectation value*** of a cross-term, that ***each*** individual cross-correlation would be ***zero***. Instead of this being an ***independence*** assumption, it therewith became an ***orthogonality*** assumption, violating the Central Limit Theorem (CLT), with serious long-term consequences.

**III)   Important differences in Fourier-space between 1D and 2D/3D data**

As discussed above, ***CCC***s depend on ***N***, the number of elements in the vectors to be correlated. For 1D data, as used in [Bershad & Rockmore 1974], a random white noise vector translates to a random white noise vector in Fourier-space. In the case of 2D data or 3D data [Van Heel & Schatz 2005] the noise is also white in a (Cartesian) 2D or 3D Fourier-space. However, we are interested in data as function of an *isotropic* spatial frequency, irrespective of the orientation of the sample within the (Cartesian) 2D image or 3D volume. We are thus looking at the data in terms of their distance ***R*** to the origin: in *rings* in 2D Fourier-space (**Fig 2**), or in *shells* in 3D Fourier-space. In these cases, the number of pixels/voxels ***N*** in the Fourier-space rings/shells varies as function of ***R*** or $R^2$ respectively (**Fig 2**). Note that the presence of symmetries in the data reduces the number of independent sampling points $N_R$ in a ring/shell.

The Shannon-Nyquist sampling theorem requires that the sampling frequency for one-dimensional (1D) data is higher than twice the maximum bandwidth ***B***. For two-dimensional (2D) and three-dimensional (3D) data, in a Cartesian co-ordinate system, however, the corresponding maximum bandwidths in the ***x***-, the ***y***-, and ***z***-directions ($B_x$, $B_y$, $B_z$), are not the bandwidths we are interested in primarily. The maximum *isotropic* bandwidth, in the 2D case, has a ring shape (up to touching the edges of the square 2D Cartesian Fourier-space). The Shannon-Nyquist sampling rule – in a strict isotropic interpretation – therefore demands that the rest of this Cartesian Fourier-space remains empty (the area beyond the disk-shaped $B_r$ bandwidth). No information may be present in the corners of this Fourier-space since that would make the orientation of an object in the Real-space image anisotropically influence its representation. Thus, in the Cartesian Fourier transform of 2D images, at least ~22% of the outer area must remain empty in order to comply with this *isotropic* 2D sampling rule. In the 3D case, the maximum *isotropic* bandwidth has the shape of a sphere inscribing the cubic 3D Cartesian Fourier-space. Here also, we require that the Fourier-space stays empty beyond the $B_r$ bandwidth sphere: at least ~48% of that sampling cube (the corners) must remain empty. We have not yet seen this issue discussed in the literature.

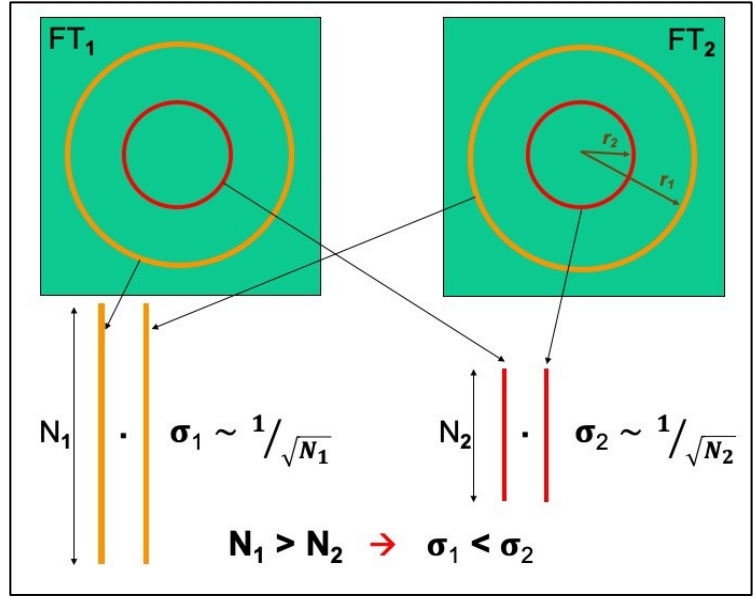

**Figure 2**: In a Cartesian 2D (or 3D) Fourier-space, the sampling is in the *X*, *Y*, (and *Z*,) directions. This Cartesian sampling scheme is not uniform as function of spatial frequency *r*. Therefore, the sampling rings far from the origin (high-frequency data components) have many more sampling points $N_1$ than the low-frequency $N_2$ sampling rings closer to the origin. This issue is of crucial importance in understanding the noise and signal contributions in spatial-frequency dependent metrics. Close to the origin, the number of sample points $N_r$ in a ring/shell drops to just a few. As a consequence, the inner products of signal or noise vectors have a large relative variance, which is very relevant for determining ***FSC***-resolution thresholds (see main text).

## IV) The sampling theorem is valid in Real-space AND in Fourier-space

The Shannon-Nyquist sampling theorem in Real-space requires that the power spectrum of the measurement being sampled drops to zero prior to reaching the (isotropic) Nyquist frequency. In Real-space this means that we cannot have abrupt changes in the measured densities such as a sharp mask delineating an object of interest, or just the sharp border of the Cartesian sampling space cutting away the measured densities. Any such sharp changes in Real-space will lead to the power spectrum in Fourier-space exceeding the Nyquist limits and introducing wrap-around artefacts. We used this classical Shannon-Nyquist sampling theorem explicitly in criticising the ***CCC***-to-***SNR*** formula boundary conditions in [Bershad & Rockmore 1974]. To comply with the sampling theorem, neighbouring sampling points in Real-space must necessarily be correlated.

It is, however, not enough that the power in Fourier-*space* gradually drops to zero prior to reaching the (isotropic) Nyquist frequency! In fact, the measurements in neighbouring sampling points in Fourier-space *must* also be correlated for the same reason that neighbouring sampling point in Real-space *must* be correlated. In Fourier-space the measurement will normally be sampled with the same number of sampling points as the measurements in Real-space (keyword: Fast Fourier Transform: **FFT**). From the Fourier-space measurement's perspective, the Real-space measurement is just its Fourier transform! By the very same Shannon-Nyquist sampling rule, we thus cannot allow for Real-space intensity components to exceed the original (Cartesian) sampling-space boundaries. Such overstepping of the boundaries in Real-space will also cause aliasing (wrap-around) artefacts, but now in Real-space. In terms of the Real-space measurement, the object or "area-of-interest" must thus also be contained within the central part of the sampling space and the information must be apodized to zero prior to reaching the edges of sampling space.

This sampling rule is the Fourier-space equivalent of the classical Shannon-Nyquist sampling rule requiring that all Fourier-space power must have gradually dropped to zero prior to reaching the (isotropic) Nyquist frequency. A rule of thumb in signal processing is to limit the data in 2D/3D Fourier-space to, say, ~2/3 of the Nyquist frequency. The Real-space equivalent of that requirement is that the object of interest should be contained within ~2/3 of the Real-space inner radius. The sampling theorem is thus a ***Janus-faced*** theorem with one face in Real-space and one face in Fourier-space! In both spaces there cannot be any power left anywhere close to the (isotropic) edges of the (Cartesian) sampling space in order to properly represent the data. We call this the ***Janus apodization rule***. This rule has direct links to the Gabor-Heisenberg uncertainty principle [Gabor 1946; Hsieh 2016]. In both Real-space and Fourier-space, the sampling must be fine enough to be able to sufficiently sample the smallest relevant detail, which, in turn, is related to the overall spread of the data in the conjugate space by the Gabor-Heisenberg uncertainty principle.

To illustrate the consequences: A Real-space spherical object like an 666Å-diameter icosahedral virus, sampled at $1.0Å^3$ / voxel, must best be contained within a box with edges of at least 1000Å. Thus, in Real-space, only $(4/3·π·333^3Å^3)$ of the $10^9Å^3$ voxels, may be occupied. That implies only ~15% of the Real-space voxels may contain information for the data to *not* violate the *ad hoc* ~2/3$^{rd}$ Nyquist sampling rule, and that requirement is matched by the equivalent requirement applying to the power distribution in 3D Fourier-space.

### V)     Traditional thinking about signals, noise, SNRs, and CCCs

We now focus on the classical measurement $x$ consisting of a fixed signal $s$ deteriorated by an independent additive noise $n$; their respective powers being: $s^2$ and $n^2$. The traditional "saloppe" way of thinking about this is that, because of the independence of signal and noise, the overall variance or the power $p$ of measurement $x = s + n$, is given by its square: $x^2 \approx s^2 + n^2$, assuming that the signal and the noise are independent. The ***SNR*** definition then follows equally saloppe as: $SNR = s^2/n^2$. However, this over-simplification leads us in a conceptually incorrect direction, when looking at the cross terms between signal and noise, and when discussing the concept of ***information***.

The ***SNR*** of a measurement always depends on the ***integration time*** $\tau$ used for attaining that measurement of an incoming noisy signal. Let us for simplicity first assume that we submit the incoming intensities to an integration time $\tau$ to obtain measurement $x(t)$. If we integrate the random-noise deteriorated signal over time $\tau$, the random-noise components average to a relatively lower value, while the signal, during that time $\tau$, averages to a relatively higher value. The associated $SNR = s^2/n^2$ value thus increases as function of an increased integration time $\tau$. There is no such thing as an $SNR$ value of a source without considering the integration time applied for collecting the vector $x(t)$. Again, the $SNR$ is a property of a specific measurement $x(t)$, collected with integration time $\tau$, and not an intrinsic property of the incoming noisy signal. (The same applies to the CCC, as stated in the introduction.)

In the single particle cryo-EM example, this integration time $\tau$ contains two elements:
**a)** the integration time $\tau$ in terms of electron counts per pixel used to collect an image, and
**b)** the number of identical particle images averaged to form the *measurement vector*.
In the latter case, if one were to average an infinite number of particle images, the noise would average out and the $SNR$ of the measurement would be infinite, through the division by the **0** noise. The $SNR$ issue will be elaborated upon in *chapter VIII,* below.

The variances and co-variances of an additive-noise-deteriorated signal are defined through the correlation values between two measurement vectors $x_1$ and $x_2$, as in:

$$\begin{aligned} p &= x_1(t) \cdot x_2(t) \\ &= \{s(t) + n_1(t)\} \cdot \{s(t) + n_2(t)\} \\ &= s^2(t) + s(t) \cdot n_2(t) + s(t) \cdot n_1(t) + n_1(t) \cdot n_2(t) \end{aligned} \quad (1).$$

There are four (cross-) terms to be considered. The first and most straight-forward term is the square of the constant signal $s(t)$; for a signal vector of length $N$ the power thus becomes proportional to $N^2 s^2$. The second term is the inner product between two *independent* noise vectors $n_1(t)$ and $n_2(t)$. The inner product of two random vectors of (sufficient) length $N$ is given by the Central Limit Theorem (*CLT*) [Wikipedia] and yields: $Nn^2$. This is because the two random vectors $n_1$ and $n_2$ have the same standard deviation $n$. For the same *CLT* reason, the cross terms between signal and noise $s(t) \cdot n_2(t)$ and $s(t) \cdot n_1(t)$ together yield an expected power of $2N \cdot s \cdot n$.

The fact that the two *independent* random noise vectors $n_1$ and $n_2$ yield a power contribution proportional to $N$ has never been disputed. However, the inner products between a signal vector $s(t)$ and an independent noise $n(t)$ have, strangely enough, mostly been taken out of the equations, using the same argumentation as used for the noise-to-noise correlations, namely, that those two vectors are *independent* or *uncorrelated*. These authors therewith **defined** any noise vector and the signal vector to be **orthogonal** (zero inner product).

**VI)** <u>**Inventory of methodological issues discussed so far**</u>

**(A)** The Shannon-Nyquist sampling theorem (in 1D) requires that a signal is sampled at a frequency higher than twice the bandwidth $B$. This implies that the power of the signal must gradually drop to **zero** *prior* to reaching the Nyquist frequency in Fourier-space.

**(B)** Sufficiently sampled band-limited data will, because of (**A**), exhibit correlations between neighbouring sampling points in Real-space.

**(C)** Two random-noise vectors will have an inner product with a standard deviation proportional to $\sqrt{N}$; where **N** is the length of those vectors. This is because the two vectors are statistically *independent* (keyword: Central Limits Theorem, ***CLT***).

**(D)** The inner product of a noise vector and a signal vector will also have a standard deviation proportional to $\sqrt{N}$ because of their independence (using the same ***CLT*** argument (**C**)). Seen from the noise vector, the signal is just another noise vector.

**(E)** For 2D/3D data, the radius **R** dependency of $N_r$ in Fourier-space must enter into the equations explicitly since that does not come naturally in a Cartesian sampling space.

**(F)** Any symmetry creates repetitions in the data. The number of effective voxels $N_r$ must be balanced against the number of symmetrical repeats.

**(G)** Isotropicity, together with the Shannon-Nyquist sampling rules, requires that for 2D and 3D data the Fourier-space power approaches zero prior to reaching the maximum inner radius $R_i$. For an isotropic data representation, the excess space must remain *empty* beyond that $R_i$ radius up into the corners of the Cartesian sampling space.

**(H)** The ***Janus apodization rule***, a direct consequence of the sampling theorem, requires that measurements in Real-space are apodized to zero towards the edges of the sampling space, just as their power in *Fourier-space* must drop to zero in time (**A**).

**VII) <u>The accumulation of flaws over time</u>**

***Resuming***: The ***CCC***-to-***SNR*** idea introduced by Bershad and Rockmore [Bershad & Rockmore 1974] included contradictory boundary conditions in Fourier-space (**A**) and Real-space (**B, H**) making the theory not applicable in practice. The orthogonality assumptions imply that all cross-terms (**C**) and (**D**) were zero. Such mathematical *a priori* assumptions, requiring abrupt changes in the data both Real-space and in Fourier-space, are unphysical while violation the Gabor-Heisenberg uncertainty principle (and the ***CLT***). Frank and Al-Ali [Frank & Al-Ali 1975] copied the ***CCC***-to-***SNR*** formula literally to the field of experimental image processing, ignoring those boundary conditions arguing that Real-space images are large (**N>>1**), thus implicitly accepting (**A-D, and H**). Moreover, Frank and Al-Ali applied the formula in 2D Fourier-space (**E**), where the number $N_r$ can be very small when close to the origin (**Fig 2**). Saxton [Saxton 1978] assumed explicitly that the *independence* of signal and noise (**D**) justified removing the cross-terms. Removing those terms, however, assumes those vectors are *orthogonal*, not independent. The argumentation (**E**) in [Saxton and Baumeister 1982] that $N_r$ is a function of **R** in 2D Fourier-space is correct. The influence of point-group symmetry (**F**) entered the discussion in [Orlova 1997] and was supported by unambiguous computational model experiments in [Van Heel & Schatz 2005], an argument mostly ignored by cryo-EM workers. An ad-hoc 0.5 FSC threshold, introduced in [Böttcher 1997], was endorsed in [Malhotra 1998]. However, that justification ignored all arguments (**A-F**) and did not provide credible arguments for that specific fixed threshold value. The new issues (**G-H**) had not yet emerged, and were obviously also violated.

The main problem in this chronological list remains the intrinsic orthogonality assumption (**D**) which was also used to deny the radial-dependency argumentation (**E**). These basic issues have been negated/ignored in many more recent methodology discussions and especially in [Rosenthal & Henderson 2003] in which paper the popular, yet incorrect, 0.143 threshold for the *FSC* was postulated. In that paper all arguments (**A-F**) were declared inappropriate using incomprehensible circular logics – no citations provided – such as: "*However, a map with or without symmetry will be equally interpretable when the FSC is the same. Any threshold criterion that depends on the number of pixels in the map is not an absolute criterion for the evaluation of resolution*".

In our supplementary materials we perform a simple model calculation in which we refute this orthogonality assumption (**D**). Our minimalistic counter example contains exactly the elements postulated in all theoretical papers, namely, that the two 3D volumes to be compared, contain the same signal but different realisations of additive random noise. The advantage of such model calculations, is that we know *a priori* what part of the correlations represent signal and what part noise. It is thus simple to assess the influence of each cross-term separately! Our model calculations show that the largest noise contributions stem from the cross-terms between the signal and the random noise, thus refuting most papers on the issue. Typically, the cross-terms are dismissed in their first mentioning, like in: "*Assuming signal and noise are uncorrelated, and for data on the same scale, the above expression may be written as follows*: …" (cited literally from: [Rosenthal & Henderson 2003]).

The fact that those cross-term contributions actually represent the ***largest*** noise contributions makes perfect sense. Since we are especially interested in the information carried by signals that are larger than the background noise, these ***s·n*** cross-terms will represent a larger noise contribution than will the pure noise-to-noise cross terms, at the critical frequencies. A sobering aspect of such methodological errors surviving for so long in the literature is that second- and third-generation methods appear in which errors have accumulated with time. An example is the recent trend towards claiming reproducible resolution levels very close to the Nyquist frequency. Such under-sampling of the data adds a violation of the sampling theorem (**G**), while using the *FSC* metric beyond its validity range, and using a refuted 0.143 threshold, which itself violates all arguments (**A-F**) listed above, etc.

### VIII) <u>Your Signal-to-Noise Ratio (SNR) may not be what you think it is</u>

Having shown that the formula used for estimating *SNR* values from *CCC* measurements is wrong, the question arises whether the formula can be corrected! Unfortunately, that is impossible because of the fundamentally different nature of the *CCC* and *SNR* metrics. The *CCC* can always be measured, the *SNR*, in contrast, is a positive-definite target that is not measurable, failing the exact *a priori* knowledge of the original signal. The historical motivation for deriving the **SNR** from the **CCC** stems from the general belief that **SNR**s are universal information metrics, a reputation we will here put under scrutiny.

The **SNR**, as a *quality metric*, can only make some very limited sense within one experiment with otherwise constant experimental conditions. It cannot serve for comparing the results of independently conducted experiments. As a simple counter example: if one bins a normal noisy image of 4096x4096 pixels by averaging 2x2 pixels into one, the **SNR** of the resulting 2048x2048 pixel image will typically increase significantly! Normal images contain strong low-frequency signal components, and significant high-frequency noise components. Thus, by reducing the number of pixels through binning, the **SNR** of the image will increase even though this will necessarily also cut out any high-resolution signal information present in the data. This simple example illustrates that the **SNR** is not a useful metric for characterising the information content of an image. In other words, the often-seen generic statements like: "*Our algorithm X works well for images with a SNR level above M*", have no absolute meaning. They cannot be used to support claims that algorithm *X* is better than algorithm *Y* because it still works at lower **SNR** levels [Wikipedia SNR].

The **SNR** can also be defined in Fourier-space in the form of a ***Spectral SNR (SSNR)*** [Unser 1987]. In Fourier-space the arguments for a frequency-dependent **SSNR** are significantly more favourable than for a Real-space **SNR**. When a Real-space image is binned from 4096x4096 pixels to 2048x2048 pixels, for example, the central part of the 2D Fourier transform of image remains virtually unchanged for both the signal and the noise part of the data and therefore the **SSNR** at low frequency also remains unchanged. The low-frequency rings used for **FSC** calculations (**Fig 2**) also remain constant, a reason why the **FSC**s are also stable for sufficiently sampled data. However, even when we focus on the **SSNR**, we cannot compare the results of different experiments on the same scale. For example, when using the same-sized images in Real-space, but gradually reducing our ***area of interest*** by masking, the **SSNR** in Fourier-space will increase gradually! This is because of the increasing correlation levels in Fourier-space. In its extreme limit, the ***area of interest*** will be a single point in the centre of the Real-space image and its Fourier transform will therefore be a constant everywhere in Fourier-space. The **SSNR** will then become infinite for all frequencies *f*, whereas any information metric should then asymptotically yield zero. Bottom line: **SSNR**s are *not* information metrics, not in Fourier-space and not in Real-space!

IX) <u>**Information is maybe also not what you think it is**</u>

The Signal-to-Noise ratio (**SNR**) concept originated primarily in electrical engineering and was designed to quantify how well a *channel* (a *telegraph line*, or a *telephone line*) transports a known input signal to its output, in spite of noise sources deteriorating that signal underway. In other words, the popular **SNR** is associated with the ***loss*** of the information when a ***known*** signal is transmitted through a channel. The fact that the **SNR** is a positive function already indicates that one assumes *a-priori* that there ***is*** a signal at the input of the channel. In data-harvesting, the story is different: we are receiving very noisy measurements at the end of the channel and we wonder whether there is some systematic signal hidden in that noise. We are interested in collecting information we knew nothing about to start with. We thus rather need metrics that measure how much information has been *harvested* from a source, not how much information has been *lost* since a known message left the source.

Real-space *SNR*s may be useful *ad-hoc* metrics in electrical engineering to describe the transport of known information from A to B, but they cannot be used to quantify small signals emerging from a background of noise. Even in the absence of a signal, the outcome of any *SNR* measurement must – per definition – be positive! Information theory was developed from that same perspective: how much of the *known* signal is *lost* by transporting it through *a noisy channel*. As per Shannon's famous rule of thumb [Shannon 1948], the information capacity of a channel is assumed to be: $InfC \approx B \cdot log_2(1 + s^2/n^2)$.

Shannon himself defines the white additive noise to be independent of the known signal in terms of *noise power* $n^2$ being independent of the *signal power* $s^2$. We note that defining the *noise power* to be independent of the *signal power* is a very different from defining the *noise n* to be independent of the *signal s*. This is because the noise-associated power components in a recorded intensity, will contain *three cross-terms*, as discussed above (*chapter V*). Of these cross-terms, the most important cross-terms are those between signal and noise: $s \cdot n_1$ and $s \cdot n_2$ (see appendix). Shannon's choice *defines* the signal versus noise cross-terms to be non-existent from the start. That definition must be seen in the context of the deterministic physics of around the turn of the 19th to the 20th century when the attitude was one of, we know it all and don't want to lose what we know. It is however, not compatible with more modern physics encompassing wave-particle duality from of beginning of the 20th century, where the square of the wave function determines the probability of finding an arriving photon or electron.

We are now facing a serious problem: the metric everybody associates with *information* and *information collection*, the *SNR*, is a definite positive metric that can actually not be used to measure *new* incoming information. Since this metric is defined as a positive entity, that makes it impossible for it to *not* collect information, even if there is none to be recorded. The *CCC*, in contrast, can always be recorded and it will naturally oscillate around the zero value in the absence of a signal. When deriving the *SNR*, through the experimental *CCC-to-SNR* formula of [Frank & Al-Ali 1975], the *SNR* is then also forced to oscillate around the zero mark, in violation with the very definition of the *SNR*. The *SNR* can thus– at best – be used to assess the *loss* of information. As a consequence, Shannon's information measure also can also only – *at best* – be used to define the *loss* of information, as per the Shannon-Hartley theorem on the channel capacity [Wikipedia Shannon]. How to resolve this conundrum? The solution is to start all over again, and create a new information metric based on the measurable and well-behaved *CCC* rather than on the evasive *SNR*. We can thus create a metric that covers both aspects of signal processing: measure the information lost in a noisy channel, or measure new information harvested from a noisy source.

# X) What if you don't *a priori* know the signal we are looking for?

Let us assume we want to collect a signal/message we know virtually nothing about prior to the experiment. Say, we purified a small protein and prepared it on an EM grid which we then imaged in an electron microscope. We have no idea whether the small protein will actually stick to the grid and what it will look like. We are thus essentially entering the territory of SETI logics (Search for Extra Terrestrial Intelligence). To grab our attention, the extra-terrestrials could choose to modulate some strong radiation source with a signal, which they would then probably repeat to emphasize that the message is intentional. We humans would then be trying to make sense out of the noisy measurements by searching for multiple copies of the message using correlation techniques [Harp 2018]. Periodic repeats of a signal will especially catch our attention (see for example [Amiri 2020]). Recognising a protein in a cryo-EM sample is probably simpler than trying to understand extra-terrestrial messages: in the case of a small protein we already have a good *a priori* idea of what a small protein would looks like in the microscope. However, trusting too much on what you think the answer will be is by itself risk-prone, as we will discuss below.

An important first parameter for data collection is the integration time $\tau$ within the transducer: each pixel of a camera – a 2D image transducer – is an independent 1D signal transducer. This recording phase is then followed by a detection-phase and an alignment-phase (also known as ***phasing***) of the received measurements. At this time, we still don't know whether we have recorded some useful information in the noisy data. One can detect the presence of messages/objects in the incoming noisy measurements by detecting peaks in the local modulation or the local variance of the incoming noisy measurements [Van Heel 1982, Burgess 1999], or using correlation functions like: cross-correlations [Saxton 1976, Van Heel 1992]; auto correlations [Harp 2018]; or triple-correlations [Kam & Gafni 1985].

Having collected what we think is useful information, we then reach a fundamental aspect of collecting unknown information: the apparently relevant new information needs to be ***validated***. Control experiments running parallel to the results experiment are of fundamental importance to verify that genuine information has indeed been collected. When authors work towards a preconceived result, science becomes very tricky. We call attention to the exchange in PNAS on cryo-EM structures of the HIV outer-membrane trimer, a primary drug target for the prevention of HIV infections. Mao, Sodrosky, and co-workers [Mao 2013] had published papers on that structure that were questionable given their processing approach. Their papers were refuted in three PNAS papers [Subramaniam 2013, Henderson 2013, Van Heel 2013]. Mao and co-workers appeared to have been working towards a preconceived answer ("*Looking for Einstein in random white noise*. The criticised work has since been superseded by a number of reliable high-resolution structures by others, but the contested papers have still not been withdrawn. This controversy illustrates that it remains difficult to prove a falsehood, especially when the underlying data is not available. When looking for something specific in random noise, one will always find matches. These techniques must be used conscientiously and be properly validated. Objective metrics help defining what information has been collected and help prevent misuse. As an introduction to our new metrics, we first discuss the classical Fourier-space correlation metrics.

## XI) The FRC and FSC metrics

The *CCC* has always been a very popular metric in signal and image-processing, in spite of its disadvantage of being indiscriminative in the case of 2D and 3D data processing. Fourier-space versions of the *CCC* like the *FRC*/*FSC* were proposed for that reason: they are much more detailed than the Real-space *CCC*, since they are calculated for each resolution interval separately (**Fig 2**). The principle behind those Fourier-space metrics is the same as for the *CCC*: to compare two signals by cross correlation. The *FRC*/*FSC* metrics have become very popular in structural biology, but they are general metrics that can be applied everywhere 2D or 3D data is being collected or compared. They are now proliferating to all fields of 2D image science and 3D tomography. The *FRC* was first used to compare the results of negative-stain EM with the results of X-ray crystallography of the same molecule [van Heel 1982]. The *FSC* was first used for comparing different 3D-reconstruction algorithms [Harauz 1986]. The *FRC*/*FSC* is a cross-correlation coefficient, in which the cross correlation is normalised by the square-root of the power in the corresponding rings/shells in Fourier-space (**Fig 2**). The *FSC* is defined as:

$$FSC(r_i) = \frac{\sum_{r \in r_i} F_1(r) \cdot F_2^*(r)}{\sqrt{\sum_{r \in r_i} |F_1(r)|^2 \cdot \sum_{r \in r_i} |F_2(r)|^2}} \qquad (2).$$

Although the product $\{F_1(r) \cdot F_2^*(r)\}$ yields a complex value, its integral over each shell/ring gives a real result, of the Hermitian symmetry of the data. The *FSC* is thus a real-valued metric, fully equivalent to the Pearson cross-correlation coefficient. When the two volumes to be compared are identical, the *FSC* yields a value of **+1**, over all spatial frequencies. When, in contrast, one volume is the negative version of the other, the *FSC* comparison yields a **-1** over all spatial frequencies. The general behaviour of this metric has been described elsewhere in extenso (cf. [Van Heel & Schatz 2005]).

The *FRC*/*FSC* has historically been used mainly as a metric for resolution assessment. But for what kind of resolution? We distinguish two main types of resolution. Firstly, there is the *instrumental resolution* of an imaging device like a light microscope, which would be determined by the physical properties of the instrument such as the numerical aperture of the lens and the wavelength of the light (electrons) used [Goodman 2004]. The instrumental resolution will ideally have a constant value over the whole object plane (*isotropic* imaging). The instrumental resolution is a characteristic of the instrument, that is valid even when the device is switched off and safely stored in a cupboard. In the **1D** case one would speak of, say, the bandwidth of an audio amplifier, but the whole idea is the same: defining the properties of the linear system, independent of the specific signals it transfers.

The *FRC*/*FSC* type of *resolution metrics*, in contrast, are primarily intended to assess the resolution actually achieved for a given sample: we call those the *results resolution*. They focus on the object we image and can refer to one specific data-collection experiment. If one here forgets to switch on the illumination, the results resolution will be very bad indeed.

Since the objects of our attention will typically be smaller than the size of the object plane, the results-resolution will not be isotropic over the instrument's field of view. The object plane beyond the object of interest, may even be masked off. All **FRC**/**FSC** results resolutions are thus *de facto* **local resolutions** when compared to an isotropic **instrumental resolution**. We tend to nevertheless call the **FRC**/**FSC** results-resolution of an area containing the *full* object of interest, a **global resolution**. The **FRC**/**FSC** results resolution may be limited by the instrument's maximum resolution, but in most cases the **results resolution** is more likely to be limited by, say, the radiation sensitivity of a decaying sample. Moreover, the **results resolution** can be compromised locally within the object, such as a full-chest tomogram, which may be blurred locally by the movement of a beating heart.

When we gradually mask off an area in Real-space to focus on a smaller area within the object of interest, the number of independent voxels in Fourier-space will be reduced due to the increasing influence of the convolution with the Fourier transform of the decreasing Real-space mask. The smaller number of independent voxels in each shell, $N_r$, must be properly be taken into account as was done for the ½-bit threshold criterion [Van Heel & Schatz 2005; 2017]. Similarly, when dealing with a structure with a high degree of symmetry, that symmetry is also reflected in information duplications in Fourier-space. The influence of these effects enters through the ½-bit threshold function, which is a function of the effective number of voxels $N_R$ at each radius (**Fig 2-3**).

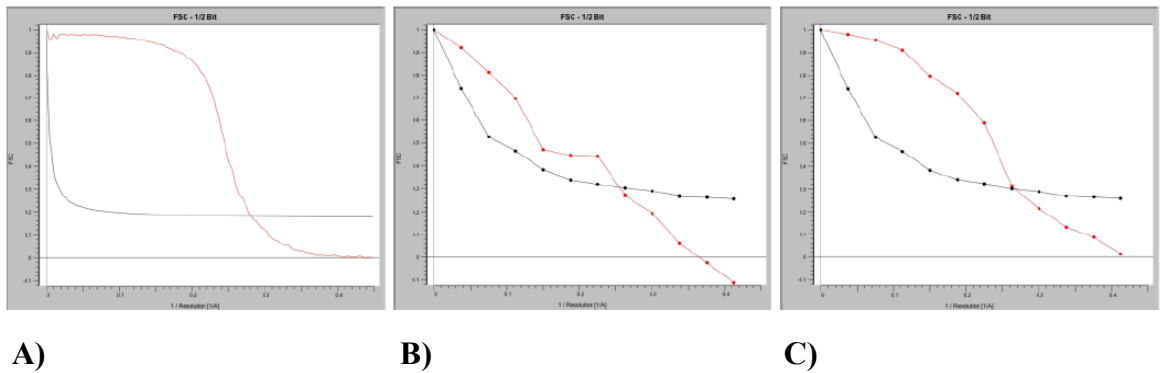

**A)**          **B)**          **C)**

<u>**Figure 3**</u>: **The *FSC* used globally and locally.**
The **FSC**, in combination with the ½-bit threshold function [Van Heel & Schatz 2005], make global and local resolution values comparable on the same scale. All three curves show a ½-bit cut-off at ~1/0.27=3.7Å, although the threshold levels differ for the global **FSC** (A) and the local **FSC**s (B-C) (from [Van Heel & Schatz 2017]).
**A) FSC** calculated for two full 3D volumes ($360^3$ voxels, 0.6 Gaussian mask)
**B) FSC** for two corresponding 3D sub-volumes ($24^3$ voxels, 0.6 Gaussian mask) cut out from the volumes used in A).
**C) FSC** calculated for two other corresponding 3D sub-volumes (again: $24^3$ voxels, 0.6 Gaussian mask) cut from the volumes used in A).

## XII) The new FRI and FSI metrics

When the *FSC*/*FRC* curves get very close to their **-1** or their **+1** limits, they no longer reflect the similarity level between the two measurements proportionally. A 1% difference in value between an *FSC*($r_i$) of **0.99**, and one of **0.999**, does not correctly reflect their huge significance difference. To compensate for this disproportionality, we apply a Fisher transform [Fisher 1915, Wikipedia Fisher], to the *FSC* or *FRC*. The resulting function we call the Fourier-Shell-Information *FSI*, or Fourier-Ring-Information *FRI* respectively:

$$FSI(r_i) = K \cdot log_2 \left\{ \frac{1 + FSC(r_i)}{1 - FSC(r_i)} \right\} \quad (3),$$

(see: **Figs 4,5**). The constant *K* in this equation is a proportionality/calibration constant with a real physical meaning, equivalent to the bandwidth *B* in the Shannon-Hartley channel information capacity ($InfC = B \cdot log_2\{1 + s^2/n^2\}$). The constant *K* will be discussed in more detail below. An alternative form of equation (**3**), emphasising the similarity to Real-space Shannon-Hartley channel information capacity is:

$$FSI(r_i) = K \cdot log_2 \left\{ 1 + \frac{2 \cdot FSC(r_i)}{1 - FSC(r_i)} \right\} \quad (4).$$

The *FSI*($r_i$) information metric is closely related to the *FSC*($r_i$) at the same radius, but is now measured directly in ***bits***, thereby eliminating the need to define a separate threshold curve. Note that these *FRI* and *FSI* information metrics are, like the *FSC* and *FRC*, subject to the rules and boundary conditions listed above, including the ***Janus apodization rule***, requiring the power in the data to drop off to zero towards the (isotropic) edges of the sampling space, not only in Fourier-space but also in Real-space. Let us now scrutinize the behaviour of the *FSI*. First, in the limit of the *FSC* approaching the zero mark, the *FSI* also goes to zero (because for |x| <<1, log(1+x) ≈ x). This was also the case for the Shannon-Hartley information capacity: when the *SNR* approaches the zero mark, the information transfer approaches zero. However, since the *SNR* is, per definition, always positive, the Shannon-Hartley information can only approach zero from the positive side.

For small *FSC* oscillations around zero, the *FSI* becomes directly proportional to the *FSC*. Small correlations are thus also directly transferred to the ***information*** level. For very high levels of correlations, both in positive and negative sense, the influence becomes logarithmic. Let us, for example, fill in the value **0.99** for the *FSC*; equation (**3**) will then yield a value of ~ **7**; for an *FSC* value of **-0.99** the answer will be ~ **-7**! Similarly, if we try the values **0.999**, and **-0.999**, the answers will be ~ **10** and ~ -10, respectively. Just to illustrate the import of correlation values close to **+1** and **-1**: for *FSC*s **0.9999** and **-0.9999** the corresponding *FSI*s are: ~ **+13** and ~ **-13**, respectively. This symmetry property of the *FSI* inherited from the *FSC*, is completely missing within the Shannon-Hartley information approach. A comparison between results of *FSC* and *FSI*-type metrics is made in (**Fig 5**).

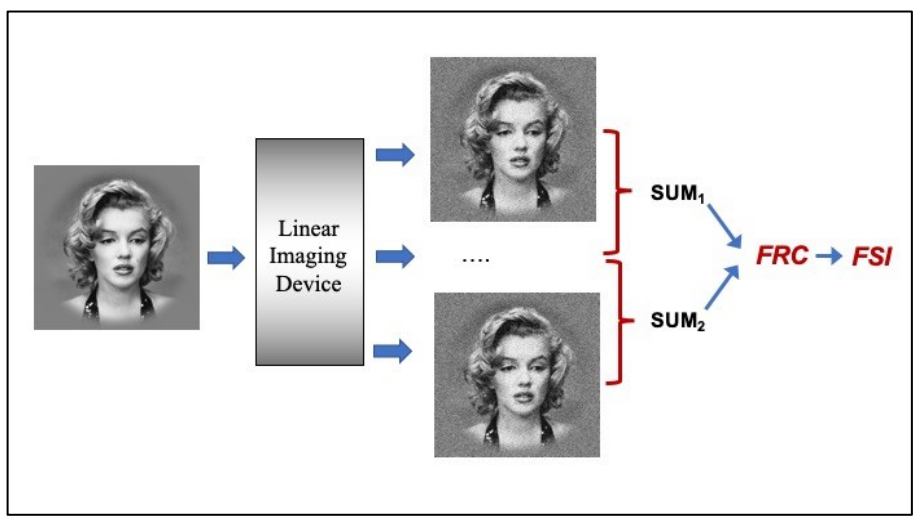

**Figure 4: Information harvesting in a *Linear System*.**
In linear systems, given a ***deterministically known object*** (here a portrait of Marylin Monroe by Richard Avedon), the ***output image*** can be predicted, provided we know the transfer properties of the linear system. Associated with this Gedankenexperiment is – in the absence of noise – the concept of the ***instrumental resolution***. It is, of course, unphysical to ever have a deterministic knowledge of an object because all information we have on the object stems from the ***noisy output images*** we have collected previously through the noisy linear imaging system. We can accumulate all images collected in an experiment into two half-dataset sums to then compare them in Fourier-space by Fourier Ring Correlation: ***FRC*** (**Fig 2**). The ***FRC*** increases with an increasing number of images being summed but it soon reaches a maximum plateau at the level **1**. More useful is to express the harvested ***information*** operating on a logarithmic scale, using the proposed Fourier Ring Information ***FRI*** metric.

To illustrate the procedures, we use an arbitrary, large experimental dataset resulting from standard cryo-EM procedures such as explained in [Afanasyev 2017]. The importance of having a large experimental dataset is that we can split it into smaller sub-sets. As long as a dataset is in a noise-limited regime, a doubling of the size of that dataset leads to a doubling of the amount of information collected at that resolution level! At resolution levels where we already have collected much information the increase in information harvested becomes logarithmic and the doubling of the data set thus leads to a linear information increase. From a large dataset with ~36000 aligned molecular images, we randomly extracted four different sub-groups, containing: 4500, 9000, 18000, and 36000 images, respectively. The ***FSC*** and ***FSI*** results from each of these four experimental datasets are shown in the same graphs, to the same scale (**Fig 5**). Different methods/algorithms discussed in this paper were used to generate the different plots (**Fig 5B-C-D**).

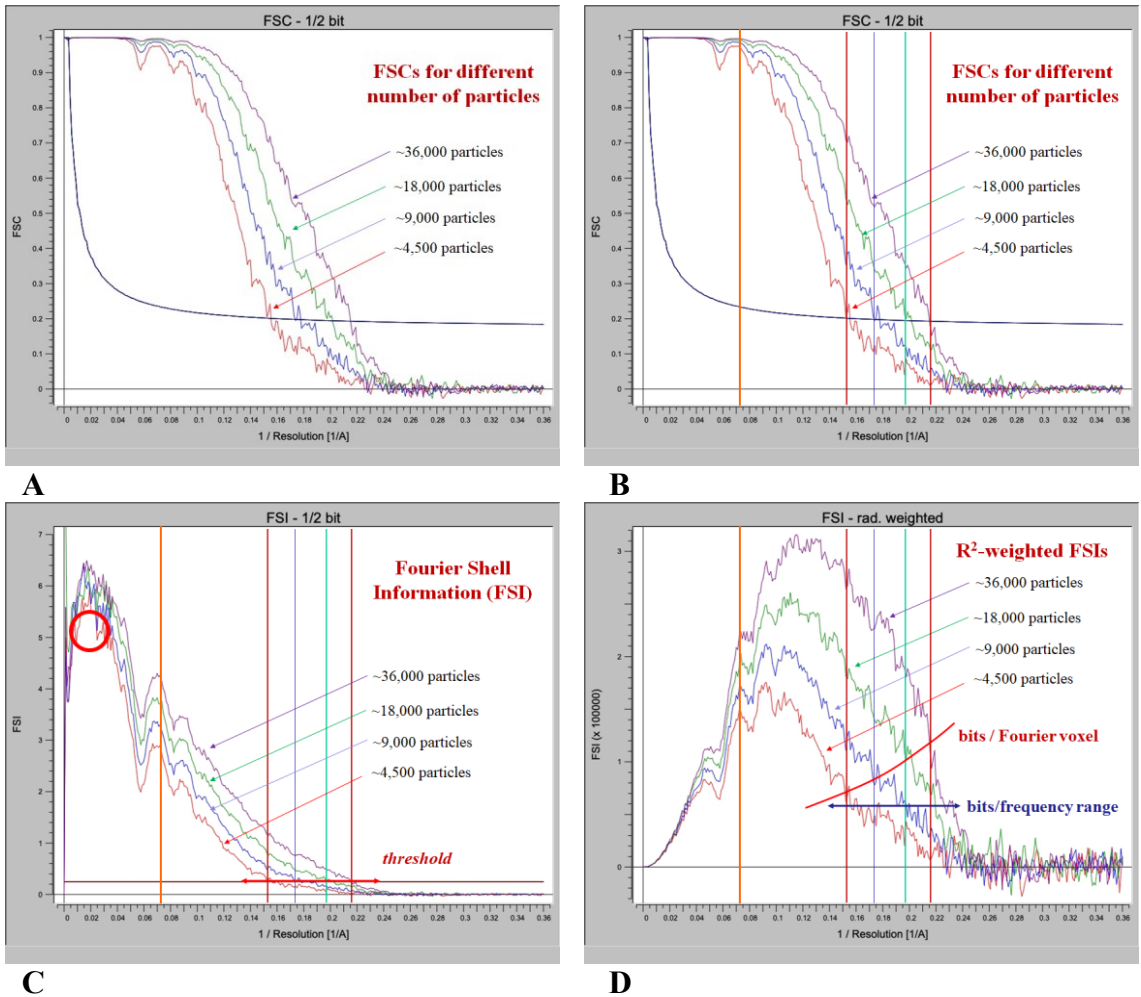

**Figure 5**: **Behaviour of the *FSC*, the *FSI* and the *FSI*$_r$, as function of dataset size.**
**A)** The *FSC* curve starts off close to the maximum value of **1** and gradually drops to a low value oscillating around the zero mark. Oscillations close to the origin are due to the small number of voxels in that area (**Fig 2**). A threshold curve indicates where enough data was collected for a reliable interpretation (the ½-bit *resolution threshold*).
**B)** The *FSC curve*s for four different groups of the dataset (1/8$^{th}$; 1/4$^{th}$; 1/2; and full dataset), calculated, with explanatory markings. The orange line, shows maxima of the FSCs very close together, close to the "**1**" maximum. The four coloured vertical lines assist in the mutual comparison of the *FSC*, the *FSI* and *FSI*$_r$ metrics resolution *threshold*s.
**C)** The *FSI* curve starts at the zero and increases gradually. The *FSI* at low resolution, however, still gives a too strong representation of the poorly defined low-frequency data (red circle).
**D)** The r$^2$-weighted FSI (*FSI*$_r$) gives the best representation of the collected information as function of spatial frequency. The brunt of the collected information is here seen in the mid-spatial frequencies, a frequency range where the amount of information collected increases rapidly with an increasing size of the dataset (see main text).

The four *FSC* curves in (**Fig 5A-B**) are all visually touching the **1** maximum, at around the 1/5$^{th}$ of the Nyquist frequency, marked with a vertical orange bar (**Fig 5B**). The four *FSI* curves (**Fig 5C**) are already showing a 30% vertical variation at 1/5th Nyquist. Moreover, for the **FSC**, the resolution threshold idea applies only in combination with the ½-bit threshold curve. In the case of the four **FSI** curves, however, the curves themselves reflect the information level achieved at each frequency. Any horizontal line will have an information content interpretation (**Fig 5C**). Note that the straight-line threshold (**Fig 5C**) on the four **FSI** curves closely match the ½-bit threshold curve crossing points on the corresponding **FSC** plots. On the low-resolution end, close to the origin, the **FSC** already oscillates away from the expected maximum value of **1** due to the small number of sampling points (**Fig 5A-B**). The few very low-frequency components hold very little information. Those contributions are better reflected in the **FSI** curves which show a real drop in the information at the low-resolution information level. The bare **FSI**, however, directly derived from the **FSC**, does not *sufficiently* reflect the low information content close to the origin.

### XIII) Radially weighting the FRI and FSI metrics

The **FRC/FSC** curves are normalised per ring/shell in Fourier-space to a value between minus and plus unity. However, the number of sampling points per ring/shell in these metrics will vary proportional to $r_i$ or $r_i^2$. The farther away from the origin, the more information each ring/shell can contain (**Fig 2**) and that information capacity needs to be weighted correctly in information metrics like the **FRI/FSI**. Other factors like the point-group symmetry and the limited spatial extent of the objects, are also needed to define the ½-bit threshold curve [Van Heel & Schatz 2005]. In the 2D and 3D cases, we thus need to weight the information contributions of the **FRI/FSI** as function of radius: the **FRI** must be weighted by the number of sampling points at each *ring* $r_i$, and the **FSI** must be weighed by number of sampling points contained in each Fourier-space *shell* (proportional to $r_i^2$). These versions of the metrics we call the $FRI_r$ and the $FSI_r$, respectively. An example of the $r_i^2$-weighted $FSI_r$, metric is given in (**Fig 5D**). Note that the $FRI_r$ and the $FSI_r$ are the metrics intended for real use: the unweighted the **FRI** and the **FSI** have been introduced more for didactical reasons. We will thus drop the *r* subscript in these and will refer to the radially weighted versions of these 2D and 3D information metrics. However, we must note here that where the *classical* ½-bit threshold in the **FRI** and the **FSI** become a straight line through the origin, or a parabolic function, respectively. What becomes more of a horizontal line with these metrics, are the total amount of information collected per frequency range.

### XIV) The constant K contains the radial weight and the κ-factor

The constant **K** is a proportionality constant with an important physical meaning comparable to the bandwidth **B** in the Shannon information capacity, albeit that **K** is rather associated with the width of the object, than with the Fourier-space bandwidth **B**. The constant **K** has a broader meaning than the analogue bandwidth **B**. The closest relative of **K** is the historical *number of degrees of freedom* introduced in [Gabor 1961] and in [Toraldo di Francia 1969]. We discussed the influence of the Fourier-space radius $r_i$ on the **FRI** and **FSI**, where, the farther from the origin, the more information a given spatial-frequency shell can carry. For 3D data (*FSI*) we can readily factorise $K_r$ as:

$$K_r \sim \kappa \cdot r_i^2 \qquad (5),$$

or, in the case of 2D data (*FRI*):

$$K_r \sim \kappa \cdot r_i \qquad (6).$$

This $r_i$- or $r_i^2$-weighting has become our standard mode of operation and we will mostly drop the *r* subscript when no explicit reference is required. Parameter $\kappa$ has the interpretation of a Real-space *filling degree*. For a three-dimensional object, with a linear dimension *D*, within a volume of linear size *L*, the filling degree is *D/L* and the value of $\kappa$ becomes:

$$\kappa \sim \left(\frac{D}{L}\right)^3 \qquad (7).$$

Similarly, in the 2D-case (images) this formula applies, but using a power of **2**. With this formulation one may first think of a cubic box of width *D*, within a three-dimensional space of width *L*, but sharp edges are not permissible neither in Real-space nor in Fourier-space since that would violate the ***Janus apodization rule***. We must rather think of a soft-edged area of linear dimension *D* containing the object of interest in a Carthesian sampling space with linear dimension *L*. The parameter $\kappa$ also has a normalisation role to play: in the limit of the object size *D* → **0**, the information content of the object also approaches zero.

The influence of *symmetry* in terms of the **½-bit *resolution threshold*** is clear [Van Heel & Schatz 2005]: in the case of a 60-fold symmetrical viral capsid structure, for example, each viral image contributes 60 individual noisy images of the ***asymmetric unit*** to the final result. In terms of the total amount information collected, as measured by the ***FSI***, however, the situation is tricky. With multiple copies of the same subunit in the final 3D results, each copy will carry the same information. One must thus scale the experiment correctly: while multiplying the collected information by the number of asymmetric units $N_{sym}$, we must simultaneously reduce the effective size of *D* to represent only one asymmetric unit. Since these effects cancel each other out, however, they are best just left out of the equations.

## XV) Local Resolution, Local Information Density, and Global Information Content

With the ***FSC*** plus ½-bit threshold combination, we have a reliable ***global*** as well as ***local*** resolution metric (***Fig 3***). Better for general use still are the ***FSI*** metrics for assessing both types of resolution as was argued above. Often the *most* useful metrics in the processing and interpretation of 3D maps, is the ***Integrated $FSI_r$***, whereby we integrate the $FSI_r$ over all relevant spatial frequencies, say, from 0.2 to 0.6 Nyquist (see information distribution over spatial frequencies of ***Fig 5D***). This ***integrated global information content*** (***GIC***) metric yields a *single value* for the information harvested on the ***global object*** and is suitable as a target optimisation parameter in iterative refinements. The ***FSC*** is often used for this purpose in combination with a specific fixed-value *FSC* threshold value. However, that is often an inappropriate combination violating the linearity of processing (see: Discussion).

Used as a *local* integrated metric, the $FSI_r$ assesses the *local information density* (*LID*) of a 3D volume as function of position. Because of the integration over the full range of relevant spatial frequencies, the *LID* is less sensitive to noise than a local *FSC* used for local resolution assessment (*Fig 3*). The *LID* can thus be calculated over much smaller local areas than the local *FSC* and, although their interpretation is similar, the *LID* is more "to the point"! If a 3D volume is (rain-bow) colour-coded by the *LID* one can directly visualize the different information levels in different parts of the object and those information levels are often directly associated with specific physical properties of that type of sample. Those physical properties could be, for example, the type of tissue in a medical tomogram or the type of rock-formation in geophysical imaging.

In cryo-EM a high level of information can be associated with rigid structural elements, whereas low *LID* values can refer to flexible parts. The *LID* in the structure of biological complexes differentiates between different areas of the structures, discriminating between protein, nucleic acid, phospholipids and glycans, contributing directly to the structural interpretation of the object. As an example, we used the hemoglobin of *Lumbricus terrestris* [Afanasyev 2017]. In the stereo-picture given in *Fig* **6A-B** the *LID* is directly associated with the structural properties of the hemoglobin. The dark-blue areas of the map are strong, high-density beta-barrel structures that form the rigid mechanical backbone of this biological structure. The red areas are primarily the more flexible parts and especially the glycosylations on outer parts of the complex. By depicting the map at different threshold levels, one can directly follow how the glycosylation extends, filling much of the remaining space between the $1/12^{th}$ subunits [Afanasyev 2017].

Our new metrics thus allow the assessment of the local information density, given the availability of two half dataset volumes. We can also apply these metrics for a cross-comparison *between* two independent volumes of identical or closely related-structures, provided the structures are properly aligned and sampled at a sufficiently high sampling rate. We call the resulting information map *local cross information density* (*LCID*).

In an ongoing study, we compare cryo-EM structures of the S-protein of the SARS-CoV-2 virus which have been deposited in the Electron Microscopy Data Bank [EMDB]. These S-protein trimer structures have mostly been released without supporting validation information such as half datasets or FSC curves, such that they cannot be verified independently. (The *L. terrestris* hemoglobin half-datasets (*Fig* **6A-B**) were also not available through the EMDB entry emd-3434, but were already in our possession). Most deposited S-protein trimer densities are under-sampled for the claimed level of resolution, since they do not adhere to the sampling rules discussed above. Most released structures have also been refined by iterative non-linear procedures to boost their 0.143-threshold based FSC resolution assessment but breaking the linearity prerequisite.

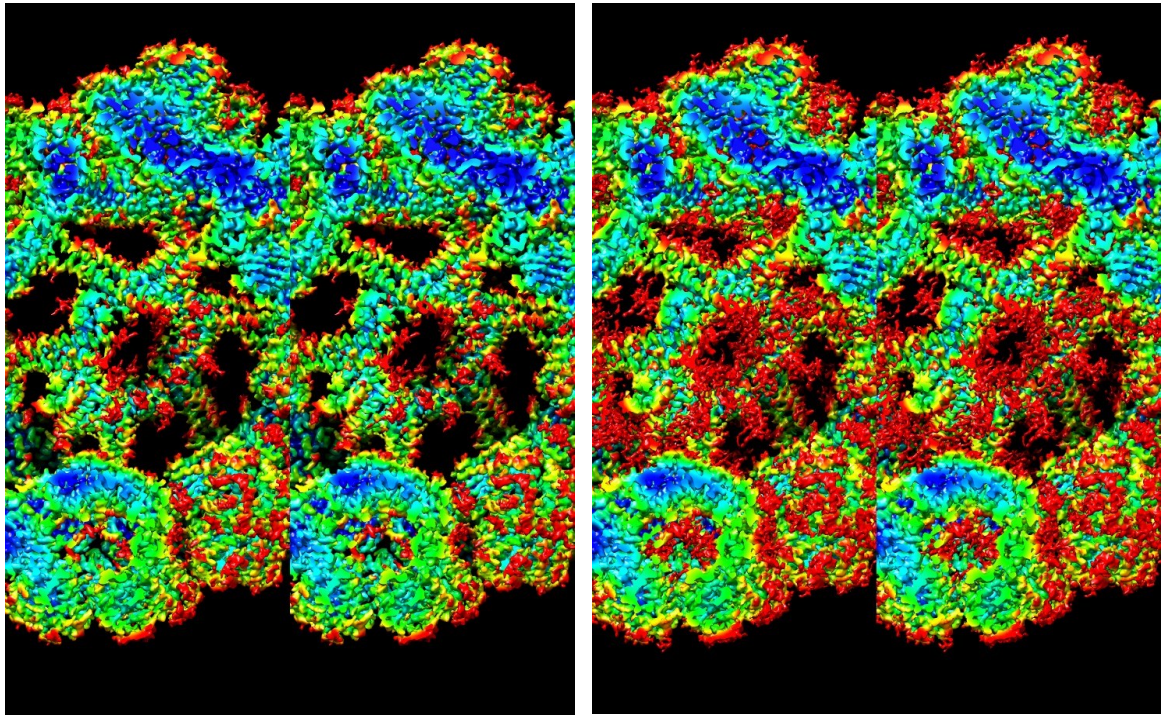

**Figure 6A-B**: **Stereo views of *L. terrestris* hemoglobin color-coded by LID.**
Using a local sub-volume (of $18^3$ voxels here), within a 3D cryo-EM $360^3$-voxels reconstruction of the hemoglobin of *Lumbricus terrestris*. The local information density (***LID***) is used to colour-code the surface of the 3.7Å cryo-EM reconstruction [Afanasyev 2017] in a rain-bow colouring scheme. The left image (**A**) is depicted using a higher threshold, thus showing less of the glycosylation (depicted mainly in red) than is seen in (**B**).

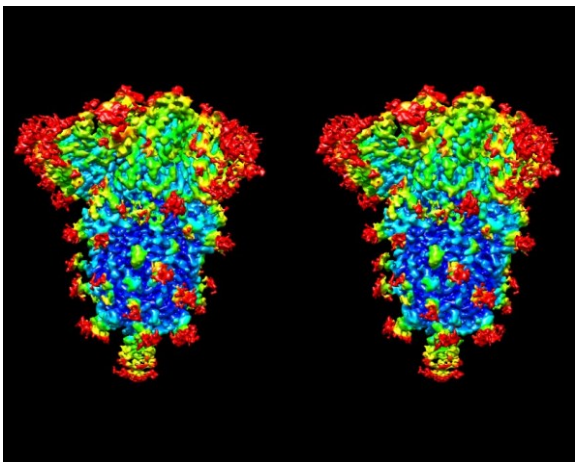 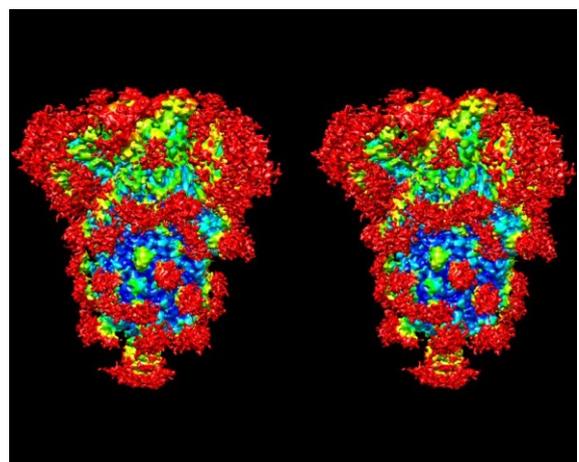

**Figure 6C-D**: **Local Cross-Information Density (LCID) views of SARS-CoV-2 spike protein.**
The map derived from EMDB entry emd-21452 [Walls 2020]) has been coloured using the ***LCID*** between two spike-structure entries (21452-11332) as described in the main text. The comparison directly reveals details of the glycosylation of this corona virus spike trimer. The level of glycan visualisation can be adapted by using different threshold levels in visualising the main map.

We have also re-sampled (and re-aligned) the structures of EMDB entries emd-21452 [Walls 2020] and emd-11332 [Xiong 2020] from 1.05Å/pixel and 1.06Å/pixel, respectively, to 0.84Å/pixel, and corrected for an additional ~0.3% difference in relative magnification. Moreover, the two structures have been brought to a matched amplitude-spectrum profile. We then calculated the ***LCID*** between these resampled maps to colour-code the emd-21452 map (**Fig 6C-D**). Of the SARS-CoV-2 S-protein structures we looked at; this was the best in terms of absence of artefacts. Others were delimited by sharp mask, having the negative densities removed, etc. The emd-21452 entry is, however, under-sampled, since with a 1.05Å/pixel sampling, no resolution level better than ~3.15Å should really be claimed.

This cross-comparison of SARS-CoV-2 spike-protein data, that we have not collected ourselves, but have downloaded from the database, emphasises the importance of correctly treating data and correctly storing such data in databases. We emphasise the importance of *not* imposing sharp masks cutting off all the background of the data, which in this case, would mean removing the glycosylation information from around the particles altogether. We have also observed the removing of negative densities from the 3D reconstructions. That violates the phase-contrast principles of the cryo-EM instrument which provides zero-average density phase information only. A further violation of the equivalence between Real-space and Fourier-space equivalency is the sharp removal of all Fourier-space power beyond a perceived resolution threshold. When different authors study the same structure, those independent studies should complement each other and augment the total amount of information harvested from different experiments. Such merging requires the studies to maintain linearity in processing throughout. Moving away from linearity in order to iteratively improve the ***perceived resolution*** of the results, may have the negative by-effect in that it represents a departure from the very concept of ***resolution*** which only exists in the context of linear imaging. The *harvesting* of as much *information* as possible, distributed over all relevant spatial frequencies, is rather the metric one needs to optimise. The direct elucidation of the extensive glycan shield [Watanabe 2020] around the SARS-CoV-2 trimeric spike-protein, illustrates the importance of maintaining consistency in our databases.

**XVI) <u>Transducer Information Efficiency (TIE) versus DQE</u>**

A further ***SNR***-related metric is the Detective Quantum Efficiency (***DQE***). This metric is used to assess transducers in medical X-ray imaging, cryo-EM, etc. The ***DQE*** is used to quantify the information lost between input and output of the transducer. It was first used in Real-space and associated with "what percentage of the quanta does my transducer actually detect". In recent years, the focus moved to a Fourier-space definition of ***DQE*** as function of spatial frequency $DQE(r_i)$. However, the historic Real-space definition remained an integral part of the definition: by defining $DQE(0)$ to be unity one tries to calibrate the height of the DQE curve in specific cases (more below). The definition of the $DQE(r_i)$ is:

$$DQE(r_i) = \frac{SNR_{OUT}(r_i)}{SNR_{IN}(r_i)} \tag{8}.$$

As we discussed extensively, the **SNR** cannot be seen as an information metric, although this Fourier-space definition is clearly better than any pure Real-space **SNR** definition (see **chapter VII**). Because the **SNR** itself is poorly defined, the **DQE** inherited that vagueness and it is thus hard to find a clear and coherent explanation of the **DQE** in the literature. The issue here that since the **SNR** is not a measurable information metric, the DQE, the quotient of two unmeasurable **SNR** assessments, is even worse off! The efficiency of medical X-ray transducers is an important public health issue, and an International Standard for DQE assessment of medical X-ray equipment has been published and updated [**DQE** Standard]. Already the introduction to this document, however, reveals the underlying, widely accepted assumption that **SNR** and *information* are almost synonymous concepts. Some quotes:
**A)** "There is general consensus in the scientific world that the Detective Quantum Efficiency (**DQE**) is the most suitable parameter for describing the imaging performance of a Digital X-ray Imaging Device. The **DQE** describes the ability of the imaging device to preserve the **SNR** from the radiation Field to the resulting digital image data."
**B)** "NOTE 1: In spite of the fact that the **DQE** is widely used to describe the performance of imaging devices, the connection between this physical parameter and the decision performance of a human observer is not yet completely understood."

In electron microscopy, the **DQE** is used to quantify different types of digital electron cameras with respect to each other [McMullan 2009; McMullan 2014]. The **DQE** measuring procedures, used in cryo-EM are similar to those used in medical X-ray imaging, and entail placing a sharp opaque edge over the transducer to block the illumination, and to then assess how the linear transfer drops off around that sharp edge. This is an error-prone linear **1D** measurement on a **2D** chip, ignoring the rest of the full area of the chip which is much better exploited in other **2D** procedures such as a camera normalisation [Afanasyev 2015]. From this **1D** measurement a Modulation Transfer Function (MTF) measurement is derived which is normalised to unity at the origin of the Fourier-space [McMullan 2009, Ji X 2019].

Following the same information-oriented thinking, as was behind the definition of the **DQE**, we now define a new Fourier-space metric, the *Transducer Information Efficiency* (**TIE**) as the quotient of the output $FRI_{OUT}$ over the input $FRI_{IN}$:

$$TIE(r_i) = \frac{FRI_{OUT}(r_i)}{FRI_{IN}(r_i)} \tag{9}$$

In contrast to the **DQE(0)**, the **TIE** at zero frequency has no special role to play. Since the **FRI** is normally even defined to have a zero value at the origin and does not carry relevant information in the data. The normalisation of the **DQE** to a unity value at the origin; mostly seen in **DQE** publications [McMullan 2009, Ji 2019], is in fact, also not inherent to its $SNR_{out}/SNR_{in}$ definition. The full publication of Ji is dedicated to the problematic aspect of correctly scaling the **DQE(0)**, primarily in medical equipment. Reasoning *ad absurdum*, when using a massive lead sheet is used as a transducer, the output $SNR_{out}$ will be zero over all frequencies. The unity $DQE$ normalisation introduced specifically for Poisson-noise limited systems then makes no sense. Another consequence of such a $DQE$ normalisation, it no longer is a universal metric to be used for all types of 2D and 3D data collection systems including, say, 3D tomography and light microscopy, etc. Indeed, the IEC 62220 standard [DQE Standard] contains statements as: "This part of IEC 62220 is not applicable to: …" among others: "X-ray imaging devices used for mammography and dental radiology", "computed tomography", etc. One more quote: "It is intended to treat some of these techniques in separate standards as has been done for other topics, for instance for speed and contrast, in IEC and ISO standard." The $TIE$ as the quotient of two $FRI$s or two $FSI$s, does not suffer from such restrictions and can serve in all cases.

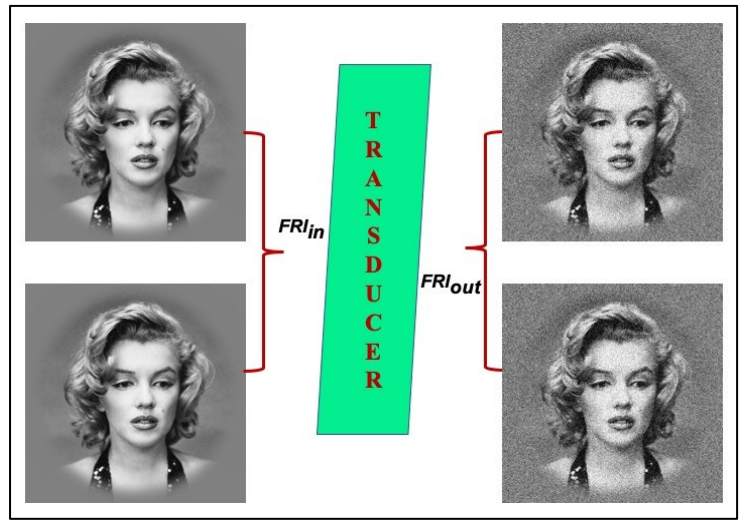

**Figure 7**: **Transducer Information Efficiency ($TIE$).**
If we know the information content of the input to a transducer (in terms of the cross information between two independently collected copies of the input signal: $FRI(r_i)$), we can compare the input information to the output information, i.e., the $FRI(r_i)$ emerging from the system. The relative information loss caused by the transducer ($TIE$) is defined as the quotient of $FRI_{OUT}(r_i)$ over $FRI_{IN}(r_i)$. For comparing different transducers, it will suffice to measure only the output information $FRI_{OUT}(r_i)$ in both cases, to find the relative $TIE(r_i)$ between the two transducers, assuming everything remains constant.

Moreover, the $DQE(0)$, is meant to calibrate the average number of quanta arriving at each pixel in Real-space. As was already discussed earlier, the very low frequency components of the $FRI$/$FSI$ metrics carry virtually no information and using the worst of those, i.e., the zero value for standardisation is not necessarily a good idea. The clearest criticism we encountered of this $DQE$ normalisation came from the manufacturers themselves [Kuijper 2015] after they found a lower $DQE$ than did their customers on the same equipment (**Fig 8**). Kuijper and colleagues commented: "… subtle differences in the implementation of the DQE measurement methods can have a significant impact on the outcome even when using the same sensor (Falcon 2)". This becomes especially clear when, the $TIE$ with radially weighted information $FRI_r$ or $FSI_r$ metrics are used (**Fig 8**) which go to zero at the origin.

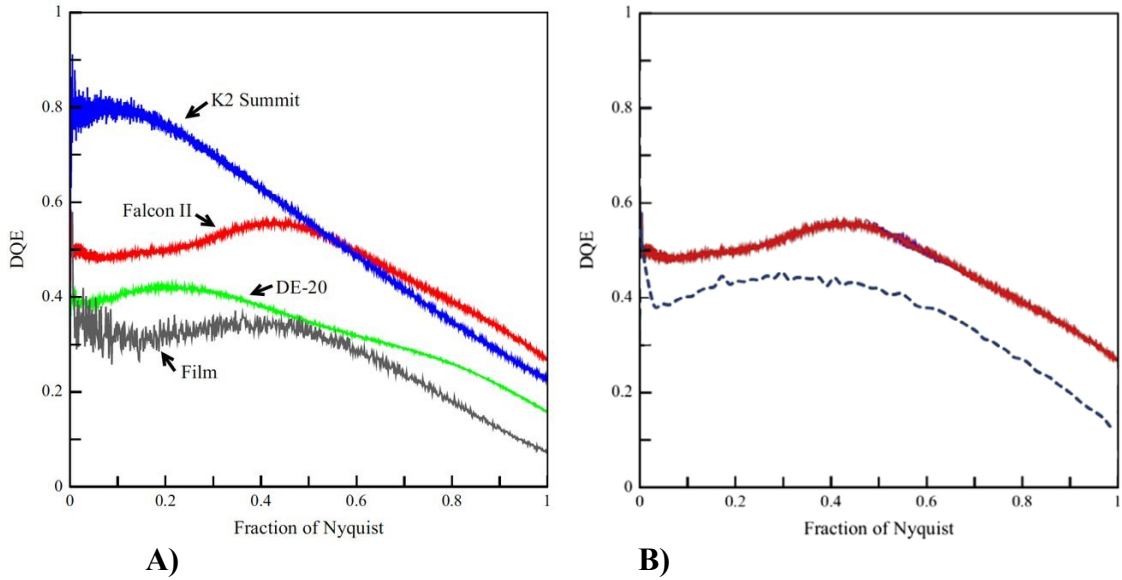

**Figure 8**: **Detective Quantum Efficiency $DQE(r_i)$ of EM transducers**.
**A)** Testing different Direct Electron transducers with each other [McMullan 2014]. Note the large discrepancy of $DQE(0)$ for these curves that have all been normalized to a unity value at the origin.
**B)** In a remarkable DQE test, the FEI camera manufacturer themselves, performing the same DQE test on the same type of Falcon II camera, come to a significantly lower DQE curve [Kuijper 2015].

The $TIE(r_i)$ will often include not only the physical transducer, but may also include additional linear imaging elements between the input object and the transducer. The transducer itself has a specific Point-Spread Function (PSF) and, with that, a specific associated Modulation Transfer Function (MTF). The MTF is normalised to unity at the origin, but the MTF is almost irrelevant because it is normalised out already in the $FRC$/$FSC$ calculation. The measurement of the transducer properties separately from the rest of the system may be problematic.

The $TIE(r_i)$ function can be used in an absolute sense but will probably often be more useful in a relative sense. With that we mean that everything else remaining the same we only need to change the actual transducer and compare the final results. The relative $TIE(r_i)$ can then be measured as:

$$TIE_{REL12}(r_i) = \frac{FRI_{OUT1}(r_i)}{FRI_{IN}(r_i)} \bigg/ \frac{FRI_{OUT2}(r_i)}{FRI_{IN}(r_i)} = \frac{FRI_{OUT1}(r_i)}{FRI_{OUT2}(r_i)} \qquad (10).$$

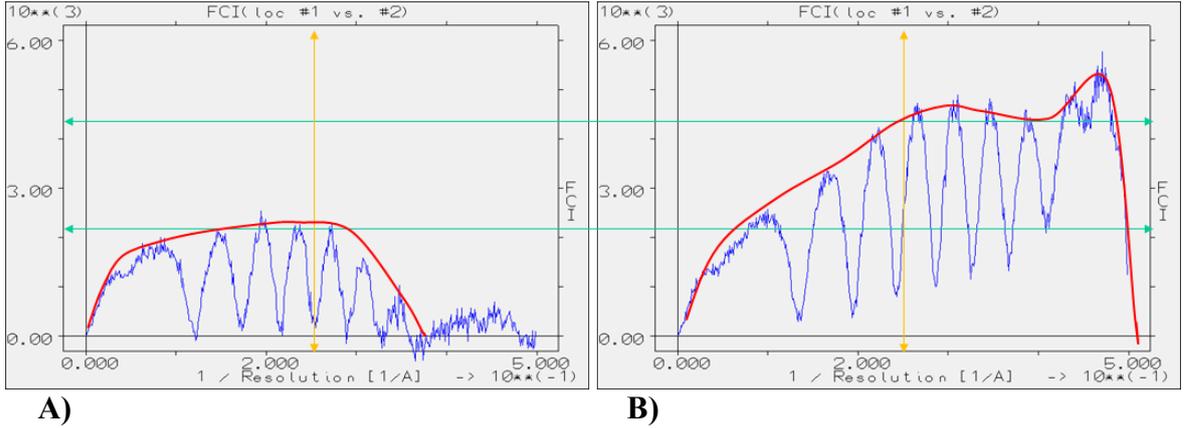

A)                                                       B)

**Figure 9**: **Fourier Ring Information $FRI_r(r_i)$ for two different cryo-EM cameras using the same test sample (Pt-Ir); both cameras were mounted on the same microscope.**
**A)** Accumulated $FRI_r(r_i)$ over 10 different measurements, measured using an "Eagle" CCD camera (FEI).
**B**) Accumulated $FRI_r(r_i)$ over 10 measurements, collected using a FEI, pre-production *Falcon II* CMOS camera.
Note that both information measurements drop to zero at the origin; the least important spatial frequency in the measurement. Interestingly, we now retrospectively assign the high bump in the $FRI_r$ (of **Fig 9B**) close to the Nyquist frequency, to the camera being capable of producing *super-resolution* results (beyond the Nyquist frequency), generating wrap-around artefacts.

The relative $TIE(r_i)$ for these two cameras results from the division of the two envelopes (marked red in **Fig 9**). In the practical test of **Fig 9**, we are evaluating the full imaging chain with two cameras mounted on the same microscope. This is not a necessity; the same sample assessed at the same magnification (and a calibrated exposure level) in different instruments remains an alternative. To just test two transducers with respect to each other, a focus, closer to Scherzer defocus (avoiding *CTF* oscillations) may be more appropriate.

### XVII) The classical 1D channel capacity revisited

In our deliberations we emphasised that the classical Shannon-Hartley channel information capacity ($InfC = B \cdot log_2\{1 + s^2/n^2\}$) is based on an inappropriate *SNR* defition. The basic model behind this metric is that of an analogue telephone-line, transferring a band-limited signal that is being deteriorated by random noise during the transfer. This is a too simple model, which leads to confusions between the integration time $\tau$ of a measurement and inverse of the bandwidth *B*. The model we prefer is that of an information packet of limited length *L*, being sent over a channel with limited bandwidth *B*. Let us first define a more comprehensive *CCC*-based metric which we call the *Packet Information Content* (*PIC*), in Real-space (*X*), as:

$$PIC_X = B \cdot \log_2 \left\{ \frac{1 + CCC_X}{1 - CCC_X} \right\} \quad (11).$$

Although quite similar to the original Shannon-Hartley channel capacity formula, this $PIC_X$ metric is not tied to the problematic positive-definite $SNR$ definition. This $PIC_X$ definition is tied to cross-correlations and compares directly to the **2D** case described by equation **(4)**. The $CCC_X$ here is the Real-space cross-correlation calculated from two different measurements of the noisy input signal, $X_1(x)$ and $X_2(x)$, collected over a specific integration time ($\tau$):

$$CCC_X = \frac{\sum X_1(x) \cdot X_2(x)}{\sqrt{\sum |X_1(x)|^2 \cdot \sum |X_2(x)|^2}} \quad (12).$$

Let us assume that the underlying Real-space analytical (noise-free) signal is $X(x)$, which is of limited length $L$, and that its Fourier transform is $F(f)$, where its power distribution is limited to a maximum bandwidth of $B$. Because of its limited bandwidth $B$, the Real-space signal can only change over distances larger than ~1/$B$, which distance in optics would be called a point-spread function (*PSF*). The product of the *PSF* and $B$ is subject to the Gabor-Heisenberg uncertainty principle [Gabor 1946; Hsieh 2016], a topic closely related to the Shannon-Nyquist sampling theorem. Over the length $L$ of the Real-space signal, we thus have available a total number of independent measurement points proportional to $L \cdot B$. This number is also known as the ***number of degrees of freedom*** of the system, a concept introduced by Gabor and others [Gabor 1961, Toraldo di Francia 1969].

The Fourier-space version of the sampling theorem we introduced above, emphasises the requirement that the sampling over the full bandwith $B$ of the signal in Fourier-space must occur over a sampling interval of better than 1/$L$, the inverse of the length $L$ of the signal in Real-space. Therefore, the number of independent measurement points available from a Fourier-space perspective is again proportional to: $B \cdot L$. All of this is still assuming we are dealing with noise-free analytical signals. Only now do we enter into the issue of actually *measuring* the incoming noisy signal over the *integration time* $\tau$. The 1D version of information capacity calculations is typically associated with functions of time (an analogue telephone-line model). That means that the issue of the integration time $\tau$ over the input signal and the mimimum sampling required by the sampling theorem 1/$B$, often become confusingly intermixed. We thus explicitly separate the two issues in our handling of the 1D case. This separation also demarcates a logical boundary between the arriving analytical waves, associated with the probability of recording intensities, and the actual counting of the arriving quanta/intensities over the integration time $\tau$. This physical measurement is then associated with counting statistics and Poisson noise, or other forms of noise. It is at this point that we measure the Real-space data and perform correlation calculation on the collected noisy vectors as per equation **(2)**. We can, of course, perform the cross-correlation calculations in Fourier-space also, in which case we first Fourier transform $X_1(x)$ to yield $F_1(f)$, and transform $X_2(x)$ to yield $F_2(f)$ prior to calculate their correlation:

$$CCC_F = \frac{\sum F_1(f) \cdot F_2(f)}{\sqrt{\sum |F_1(f)|^2 \cdot \sum |F_2(f)|^2}} \qquad (13).$$

Note that the cross-correlation coefficient in Fourier-space $CCC_F$ has a real value (as was the case for the **FRC** and the **FSC**) because $X_1(x)$ and $X_2(x)$ are real and therefore both $F_1(f)$ and $F_2(f)$ are hermitian, a property that their product inherits and while summing a hermitian pair, only the real part survives. And, based on the $CCC_F$ we can calculate the Fourier-space version of the $PIC_F$:

$$PIC_F = L \cdot \log_2 \left\{ \frac{1 + CCC_F}{1 - CCC_F} \right\} \qquad (14).$$

When the signals in Real-space and in Fourier-space are approximately uniform and well-behaved in terms of the ***Janus apodization rule*** and simultaneously remain far from fundamental limitations imposed by the uncertainty principle, the Real-space and the Fourier-space version of the **PIC** will yield identical results.

We have now repeatedly mentioned the possible confusions between the integration time $\tau$ of a measurement, and $1/B$ determining the minimum sampling frequency required to correctly catch the analytical noise-free signal we are trying to record. In case of single particles in cryo-EM, we explicitly mentioned the two aspects of that integration time $\tau$: firstly, there is the actual integration time of the arriving signal in the transducer, and secondly, the number of independent particles we average into the final measurement prior to the **CCC** measurement. When thinking in terms of a 1D signal in the time domain, this is where things start to get confusing. If the measurement comes out too noisy, the natural reaction of an electronics engineer is to put in an extra capacitor in the A/D conversion to increase the integration time $\tau$ so as to reduce the noise in the resulting measurement. However, that is exactly the wrong reaction, because such increasing of the integration time $\tau$ will suppress the natural bandwidth **B** of the signal! The correct reaction is rather to repeat the measurement multiple times, following the better-than $1/B$ sampling theorem, and to then average the independent measurements!

The information capacity of the 1D channel must be seen in this context. If the **PIC** is seen as insufficient for the purpose, we need to repeat sending packets until the underlying information is recorded to a sufficient level of redundancy; only then can we move onto record the next sequence of packets. The channel capacity will simply be the product of the individual **PIC**s times the number of packets that can be sent per second.

## XVIII) Discussion

We have not been able to trace an exact historical origin of the *SNR* but it obviously became a metric in telephony and electronics around the turn of the 19$^{th}$ to the 20$^{th}$ century. We can already see extensive use of logarithmic *SNR* forms measured in decibels (*dB*), which are useful for assessing high *SNR* situations, and which represent a clear movement towards information-based thinking. The various discussions in [Wikipedia SNR] do reflect that. In those days large-scale digital data processing was still to be invented and there thus was no feed-back from data-harvesting experiments. The flaws of the *SNR* as a practical metric remained in the background.

The *SNR* concept is deterministic in the sense that one must have absolute knowledge of the input signal one wants to transmit through a noisy communication channel in order to see how much information will still be arriving at the end of the channel. A special abbreviation was sometimes given for this hypothetical type of sample: **SKE**, for Signal-Known-Exactly [Burgess 1999]. Having absolute knowledge about anything is not a realistic premise since it requires an infinite amount of time to collect such information. We are not trying to be pedantic, just to be exact: the ideas behind the *SNR* definition were not realistic, and violate basic principles of modern physics. It is impossible to separate a signal from the noise at the output of the communication channel in real-life experiments; the *SNR* concept is thus to be used more in thought-experiments rather than used as a real-life metric.

For the *SNR* one must always make specific *a priori* assumptions about the behaviour of both the input signal and of the assumed additive noise that interacts with the signal during transmission. We have encountered various, often contradictory *a priori* assumptions such as in [Bershad & Rockmore 1974], [Frank & Al-Ali 1975], and above all in [Shannon 1948] where orthogonality is assumed in ($s^2 \cdot n^2$). In today's view of data collection, quanta carrying the information will be detected in a transducer in the course of the integration time $\tau$, leading to both the signal and the noise. From the theoretical perspective, there is neither a signal nor a noise prior to their detection, just wave functions. The probability of quanta arriving at the image transducer is given by the square of that wave function. Such fundamental principles are, never really touched upon in the context of the use of the *SNR*. The Bershad and Rockmore mode of calculating the inner product between independent measurements, is more compatible with the wave function idea, but that was then directly defined out of existence by removing the cross terms. The Shannon approach, remains entirely within a deterministic *particle world* in which all is countable and there is no correlation between the *power* of the signal and the *power* of the independent noise. This thinking is not compatible with the concepts of the wave-particle duality.

In Shannon's information theory, the prior probability of a measurement must be known beforehand. The classical example given in the literature for Shannon's information is the flipping of an ideal coin where the outcome can have only two values: heads or tails. The more elaborate example is the *a priori* knowledge that an incoming message consists of, say, the 26 letters of the English alphabet. One then needs to collect enough of the noisy signal to be sure, of which of all possible messages one has received. There are clearly many more possibilities in the latter case than in the coin-flipping case, but the prior-probability space is equally deterministic as is the heads/tails prior-probability space. Others have also pointed out (some of) the limitations of classical information theory, see, for example, the work by Deutsch and Marletto [Deutsch & Marletto 2014] on **Constructor Theory of Information**. To quote these authors on Shannon's work: "Much of Shannon's theory is about unreliable transmission and measurement, and inefficient representations, and how to compose them into more reliable and efficient ones". Quoting Deutsch and Marletto on their own work: "… receiving the message means *distinguishing* it with perfect reliability from all the other possible messages." This information theory is thus also aimed at distinguishing a message from all possible messages, which themselves must be known *a priori*. All clarifications/extensions of information theory we have seen, do not properly integrate information theory into the more down-to-earth case of receiving unknown data in a signal-processing context.

We have already mentioned that in real-life experiments it is impossible to separate the signal from the noise in a measurement. In fact, in the physical reality of data collection, the model of a "signal plus independent noise" is not only unrealistic and unnecessary, in the case of arriving quanta it also violates modern physics principles, as discussed above. For the **FRC**/**FSC**, and the new **FRI**/**FSI** metrics (as well as the **PIC** metric) we just introduced, we need not make any *a priori* assumptions about signal and noise. Under ideal circumstances the information will gradually add up in the recording instrument and the quantum noise associated with the arrival of the individual photons/electrons will slowly disappear under control the graceful central limit theorem. We ourselves have used $(\boldsymbol{s} + \boldsymbol{n})$ models explicitly in our appendix, for example. However, that was only while discussing the errors usually made in the field, in assuming that $(\boldsymbol{s} \cdot \boldsymbol{n})$ products are negligible compared to the $(\boldsymbol{n_i} \cdot \boldsymbol{n_j})$ products. With our metrics, such underlying models become superfluous. The metrics are independent of any *a priori* knowledge of the signal or the noise. In maximum-likelihood (ML) approaches, in contrast, *a priori* assumptions about the noise versus the signal behaviour are the basis of the ML analysis.

In Fourier-space, the information on the object will gradually drop-off towards the high-resolution end of the scale. By defining a correct cut-off point, a reproducible resolution label can be attached to the measurement, say, to be deposited in the data-base. Note, in this context, that the ½-bit criterion we had introduced it in our 2005 paper [Van Heel & Schatz 2005] was aimed at including the missing $\boldsymbol{s} \cdot \boldsymbol{n}$ cross-terms into the *traditional **SNR** concept*. We originally included those into the definite-positive **SNR** definition and therewith more into Shannon's information concept. We suggested the ½-bit information level as an **FSC** significance threshold which occurs when the cross-term corrected **SNR** reaches the 0.4142 value [Van Heel & Schatz 2005]. That definition, however, did not yet include our new insights of dismissing the positive-definite **SNR** altogether, and instead including the possibility of negative information. We do continue to propagate the ½-bit criterion as a significant rule-of-thumb threshold for cryo-EM data collection.

We want to emphasize that even with our re-definition of the **bit** the earlier ½-bit threshold remains essentially numerically the same! What hopefully becomes clear from our deliberations is that a high, single-valued resolution quote that all desperately try to achieve based on a reproducible *resolution threshold*, is only of secondary importance! The first paper where the *FSC* was ever mentioned [Harauz & van Heel 1986] only provided *FSC* curves, never a single cross-resolution value. What one tries to achieve is to collect as much linear information as possible from our object of interest, over a large range of spatial frequencies, to be able to usefully interpret it.

The idea of information harvesting emphasises the important fact that two independently determined maps of the same object will supplement each other and can be merged to yield an overall even better final result. Two independent submissions of the same object to the databank represent more information than each submission individually, even if they were determined to the same level of resolution. The cross-resolution between the two submissions should therefore be higher than that of each submission individually. The information contained in two independent cryo-EM volumes can therefore be pooled to form a better map, which can serve to generate an overall better structure which can be used for, say, drug development. This complementarity can best be achieved, by always processing the data within the context of *linear systems*. The new information metrics can serve to quantify and merge that complementary information.

The concept of *resolution*, either in terms of the maximum *instrumental resolution* or the best *results resolution*, exists only within the framework of *linear systems*. Linear systems operate under the established rules summarized in this paper. Iterative refinements, boosting the high-resolution data components, while targeting a maximization of a metric like a fixed-valued *FSC* threshold, will make the overall system deviate from linearity. Similarly, so called "density modification algorithms", typically inherited in some form from X-ray crystallography, are aimed at changing the measured cryo-EM density in order for atomic models to better fit the no-longer experimental densities. No matter how well the resulting densities/atomic will match each other, these procedures violate the basic linear information transfer rules. As a consequence of moving away from that theoretical basis of linearity, the outcome of independently conducted experiments become incomparable to each other, even when they are associated with the same claimed *resolution* value. One can strictly speaking no longer refer to a (*FSC*) *resolution*, simply because one has left the realm of linear-systems in which the concept of *resolution* is defined.

The new *Fourier Ring Information* metric ($FRI_r$) may also fulfil the more abstract role of defining how good an image compression algorithm performs, for example. One can calculate the cross information between the input and the output of a data compression algorithm and thus objectively measure how well the algorithm performs independent of a team of human observers who traditionally would have the final word on what is good or what is better in image data compression. Possibly more important is the assessment of *algorithms* used for 3D reconstructions in, say, medical tomographic equipment. Note that this was the very first published use of the 3D *FSC* [Harauz & van Heel 1986].

A direct consequence of our re-assessment of the meaning of the *SNR* and *Information*, is that the *DQE* must also be replaced by a more appropriate metric. Our new *Transducer Information Efficiency* ( $TIE(r_i)$ ) perfectly fulfils the expectation of a metric for assessing transducers. It can incorporate a full data collection chain and not just the direct physical transducer itself. An example of such a full chain is, for example, a digital photographic camera, including a large aperture lens, which we wish to test at different apertures and different illumination conditions. The approach can also be applied to a full chain of medical NMR or X-ray tomography equipment which we want to optimise to a best possible visualisation of small calcifications in the human brain, for example. Our approach for creating standards would typically include specific test objects like Pt-Ir samples in electron microscopy, or specific phantoms for mammography, or for medical NMR, X-ray or tomography. Using such standardised samples, one can subsequently assess the total information collection chain, including not just the physical transducer, but also assess the total amount of information at a resolution level required for a specific diagnostic purpose.

Let us now return to the Shannon's information concepts in a 1D world, to emphasise what changes and what remains the same with the introduction of our new metrics. There was nothing wrong with Shannon's ideas around coding and the loss-less retrieving of stored information; the modern world would be unrecognisable without those principles. What everybody – and that included us authors for decades – has apparently missed is that for Shannon's concepts one requires to have a total *a priori* knowledge of the information one wants to transfer. This *a priori* knowledge can be a deterministic knowledge of the signal/message itself, or of all the possible symbols a message can consist of, like the letters of the alphabet, or a perfect knowledge of all possible quantum states an atom can assume. Only in such a deterministic framework can one define the required probabilities. However, when transferring that deterministically known signal through a noisy channel, the cross terms between that signal and the random noise must be included in the deliberations.

The harvesting of new information was never central to the thinking of communication scientists in the early 20$^{th}$ century. The signal and noise property assumptions that Shannon added, in defining his channel capacity, were logical against the engineering background of the *SNR* but were unphysical in terms of modern physics. In more modern times Bershad and Rockmore [Bershad & Rockmore 1974], Frank and Al-Ali [Frank & Al-Ali 1975], Saxton [Saxton 1978], Rosenthal and Henderson [Rosenthal & Henderson 2003], and various others, did calculate the intensities as the inner-product between two different measurements of the signal plus noise. But, by directly declaring the cross-terms between the measured signal components and noise component vectors to be orthogonal, modern physics was again ignored. In our previous papers [Van Heel & Schatz 2005; 2017] we argued that including the cross terms in the *CCC*/*SNR* calculations is obligatory for a correct *SNR*-based assessment. The pre-cursor of this paper was accompanied by a list of publications that were incorrect for that reason alone [Van Heel & Schatz 2017]. We now realise, that including the cross-terms in the *SNR* calculations may have been an important step in the right direction, but it was not a sufficiently large step. The *SNR* itself, and hence the classical channel capacity must also be rejected as metrics to assess new information. We conclude that *CCC*-based metrics leave little to be desired as practical metrics that can be integrated into a practical information-science framework.

As we mentioned above, a real historic complication was that the *SNR* and *information* ideas emerged in the world of 1D time-domain processing, that is, of voice and telegraphy processing. The integration time $\tau$ in the transducer and the Real-space sampling step which by the sampling theorem needs to be smaller than 1/2 *B*, were hardly ever separated in the early literature, or even in the modern literature. This fact obscures the basic duality between the Real-space and the Fourier-space representation of the data. We emphasised the necessity to apply the sampling theorem rules in both Real-space and Fourier-space. Not only must the power of the signal die out in Fourier-space prior to reaching the Nyquist frequency, the same is true in Real-space where signal must gradually die out prior to reaching the (isotropic) border of the sampling space. The sampling in Fourier-space must thus be finer than 1/2*L* where *L* is the linear size of the signal in Real-space.

The symmetric treatment of these conjugate spaces brings us close to the thinking of pioneers like: Gabor, Hartley, Toraldo di Francia, and Heisenberg and their introduction of the ***uncertainty principle*** and the ***number of degrees of freedom***. Their thinking was, however, in terms of noise-free analytic functions and did not incorporate noise and the collection of new information. By including noise and *SNR* and *information* concepts in the processing of 1D time-domain data, the basic issues became obscured. In collecting 2D image data, the integration time $\tau$ in the transducer, takes place in an additional dimension (i.e. time), other than the X-Y spatial dimensions in which that the image data is being collected. When translating that extra dimension of the integration-time $\tau$ back into the classical domain of the assessment of 1D time signals, we explicitly included that $\tau$ dimension in a form independent of the Real-space sampling. We added that in through the averaging of different noisy packet with the same underlying signal. This form is inspired by the cryo-EM routine to average multiple noisy images of a molecule to build up the basic signal. In suggesting a successor to the concept of the channel information capacity, we thus have chosen the ***Packet Information Content*** (***PIC***), a choice which restores the symmetry of Real-space *versus* Fourier-space data and clearly separates the issue of integration time $\tau$ from the sampling step in Real-space. The packet concept also fits well into the practice of today's digital communication and also help us move away from the infinite integration limits of traditional analytical functions.

Our information metrics for 2D and 3D data, namely the ***FRI*** and ***FSI*** (equation **3**) have been proposed using correlation functions in Fourier-space, the ***FRC*** and the ***FSC***, respectively. These information metrics are proportional to ***K***, which factor is largely dependent on the size of the object in Real-space. The underlying assumption in this formulation is that the distribution of the information in Real-space is approximately constant over the area/volume of the object. In Fourier-space the information is not evenly distributed (as per Fig **2**) hence everything must be weighted correctly in Fourier-space. In the 1D case, on the other hand, we calculate the correlation/information in Real-space assuming an approximately constant bandwidth ***B*** in Fourier-space. The implicit assumption here is that the information in Fourier-space is evenly distributed in this 1D case. These are, however, in both cases, rules-of-thumb that are not carved in stone since we are operating in conjugate spaces that – overall – must contain exactly the same information.

The crux of our *local metrics* is precisely to have a *local* assessment of information content of the data such as in our example of the corona-virus spike protein, because the information is *not uniform* over the full volume of the cryo-EM reconstruction of this spike protein. The departures from a uniform information distribution are directly related to the physical/biological properties of those areas which thus directly help the interpretation of the object. In the 1D case also, the Real-space signal within the sampled measurement need not have a constant frequency bandwidth over the measurement. When recording music, for example, the frequency content of the data will, of course, change continuously with time and it becomes important to assess the information harvesting on a more local basis in the sense of the Gabor transform. In the Gabor transform [Gabor 1946, Wikipedia Gabor Transform], sounds are analysed in the time domain and the frequency domain simultaneously. However, Gabor's pioneering concepts were never intended for use in an information harvesting context. To follow the philosophy of our current paper, one would need to record the sounds twice (preferably twice simultaneously) and assess its (local) information content in terms of the correlation-based information metrics in Real-space (as per equation **11**), and/or, in Fourier-space (as per equation **14**).

Where does the influence of the transfer function enter into in this information theory discussion? We have mentioned the concept in-passing throughout his paper. The (*Phase*) *Contrast Transfer Function* (*CTF*) and linear systems are two closely tied concepts and they emerge in the first place in the definition of the properties the instrument like the maximum bandwidth *B*, or the *instrumental resolution*. Strictly speaking, every time we change the focus of a (electron) microscope, we create a new instrument with different instrumental transfer properties [Van Heel 1978; Goodman 2004]. The effect is clearly visualised in the *FRI* measurements as shown in **Fig 9**, where we have hand-drawn a red envelope curve to emulate an ideal instrument with a more constant transmission over all relevant spatial frequencies. In cryo-EM one tries to remove the influence of on the *CTF* by collecting that data over a range of different defoci to thus create a virtual instrument with an effectively continuous positive transfer function. With a sufficient number of images collected this then largely creates an almost perfect virtual instrument. Nevertheless, one can still recognise in **Fig 5** some low-frequency dips in the information transfer due to the oscillating nature of the *CTF*. (Note that the often-used *Modulation Transfer Function* (*MTF)*, the absolute value of the *CTF*, violates the linearity of the procedures).

The damping of the transfer function by the presence of an aperture or other limitation will normally (or even preferably) quench the information transfer through the optical system to zero towards the (isotropic) Nyquist frequency. This can be observed in **Fig 5D** where all four *FSI* curves start oscillating around the zero-value prior to reaching the Nyquist frequency. We here speak of "preferably" because when incoming signals are not sampled sufficiently for their bandwidth, wraparound artefacts are introduced in violation of the sampling theorem. It may thus be better to "low-pass filter" the incoming signal by a limiting aperture associated with the appropriate instrumental resolution. Whereas no information should present close to the Nyquist frequency, farther from Nyquist, when the information content is still low, a doubling of the number of measurements also doubles the information harvested (the rightmost vertical line in **Fig 5D**). Farther still from Nyquist in **Fig 5D** (leftmost vertical line), when the harvested information level is high, a doubling of the number of measurements leads to a linear increase in the level of harvested information. Both effects are a consequence of the logarithmic aspect of information collection.

A final question remains: If the *SNR* is indeed such an erroneous metric, how could it become such a dominant cornerstone of science for over a century, whereas a correct and measurable *CCC* has been available all that time? In retrospect the reason is understandable: in the limit of *CCC* → +1 the predictions derived from the *SNR* become identical to those made with the *CCC*. In other words, the predictions of the *SNR* for extremely high *SNR*-values limit approach the exact behaviour. In the practice-oriented field of electronic engineering and signalling, the routine use of the logarithmic decibel (*dB*), preceded the publication of the Shannon and Hartley concepts of information. In hindsight we can now argue that the *dB*, in the high-*SNR* regime, gives identical results to the information metrics we introduce here (ignoring proportionality constants). The problems of the *SNR* in relation to the *CCC* became very visible in the disastrous formula (*SNR = CCC / (1-CCC)*), where small oscillations of the *CCC* around zero lead to undefined negative *SNR* values. As scientists pre-programmed by a classical physics education, it also took us, the authors of this paper, too long to realise that the primary problem was not the *CCC*, and the errors made using it, but rather the *SNR* definition itself that relates only to the loss of information.

Our new information metrics can assume negative values, especially important when collecting noisy information. The new metrics allow us to properly integrate information principles into the vast field of signal-processing. We do need to accept that collecting positive bits of information is equally important as collecting negative bits. That is an essential part of a defining consistent information metric. To those wondering how negative information can be so important, you are reminded of the classical mathematical puzzle on how to extract a correct answer out of a population consisting of systematic liars and systematic truthtellers. The bottom line of the puzzle is that systematic liars provide us with just as much information as do individuals who always tell the truth.

**XIX) Conclusions**

We have argued that *SNR*s and the *Information channel capacity* are metrics aimed at minimizing the loss of *known information* during transfer; they are not useful for quantifying the harvesting of new information. New information can be collected using cross-correlation techniques where cross-correlation coefficients (*CCC*s) play a key role, but the 1970s formulas relating *CCC*s to *SNR*s, were fatally flawed. In fact, all data collection metrics implicitly or explicitly assuming that *SNR*s are reliable information metrics, must be rejected. The *SNR* and its associated information metrics, require a *deterministic knowledge* of the object of interest (the signal) prior to any experiment. Our new information metrics are based on *CCC*-type metrics in Real-space and in Fourier-space. To replace the concept of Shannon-Hartley channel capacity, we propose the Packet Information Content (*PIC*); the channel capacity follows by multiplying the *PIC* by the number of packets per second. For images and 3D volumes, the information metrics are based primarily on the *FRC* and *FSC*. The new metrics, named the Fourier Ring Information (*FRI*), and the Fourier Shell Information (*FSI*), are also measured in bits. These bits are measurable entities that can assume positive or negative values, a property inherited from the Fourier-space *FRC* and *FSC*, or from Real-space *CCC*s. These metrics oscillate around the zero mark when no signal is present. We have also introduced a new Transducer Information Efficiency (*TIE*) metric to replace Detective Quantum Efficiency (*DQE*). We demonstrated our information metrics using *SARS-CoV-2 spike structures* deposited in the EMDB databank.

## Acknowledgements

The ideas expressed here have matured over decades with significant contributions from various colleagues in different laboratories. We especially thank George Harauz for his early contributions at the Fritz-Haber Institute in Berlin. We thank Ralf Schmidt (Image Science, Berlin) for continuous support during the methods development. Ka Ho Tam performed early FRI tests when a student at Imperial College. Gijs van Duinen of FEI (now Thermo Fisher) collected the micrographs used for generating the TIE test data. Juliana Mello da Fonseca pre-processed the worm-hemoglobin data. Sayan Bhakta (of CSIR - Indian Institute of Chemical Biology, Kolkata) performed the local-resolution tests on corona virus spikes. Moritz Quincke, a summer student from Ulm University, performed under-sampling tests and proof-red the manuscript. Rodrigo Portugal and Adalberto Fazzio have been sparring-partners in information discussions. Helder Veras (LNBio, Campinas) contributed to the information plots (Fig 5). We are eternally grateful to Tim Balance, a physics student at Imperial College at the time (2011), for nicknaming our negative information *slander*.

**Note 1**: We appreciate suggestions for submission of this manuscript to specific journals. We do wish to avoid anonymous referees because of conflicts of interest in a world where "everyone" probably has made the fundamental errors we here criticize.

**Note 2**: All 2D/3D data-processing and testing in this paper was performed in the context of the IMAGIC-4D data processing / development system, running on notebook and desktop computers under the Linux, Windows, or Mac OS systems [Van Heel 1996; Van Heel 2012]. We used the UCSF Chimera program [Pettersen 2004] for displaying local information density results in stereo (in Fig 6A-D).

*Supplementary Materials*

# A simple model experiment

The gold standard used for assessing the quality of 3D reconstructions in cryo-EM is the Fourier Shell Correlation metric (***FSC***) applied between two independently determined volumes [Harauz & van Heel 1986]. The *a priori* assumption made for such comparisons in theoretical papers is that the two volumes (**Fig S1a-b**) contain identically the same 3D information (the signal ***S***, **Fig S1d-e**), but that each volume is deteriorated by a different realisation of an additive zero-mean random noise ($N_i$, **Fig S1f-g**). In our simple test, ***S*** is the structure of a giant hemoglobin [Afanasyev 2017], but that is immaterial to this model experiment. The two volumes: $A = S+N_1$ (**Fig S1a**), and $B = S+N_2$ (**Fig S1b**), are generated using the same starting volume ***S***, and adding two random-noise volumes $N_1$ and $N_2$, respectively. Important to the experiment is that we know all these components *a priori*, such that we can assess all cross-correlation components separately, which is a luxury one never has in real experiments. From the 3D Fourier transforms of these 3D densities (***A*** and ***B***; transforms denoted in italics) all correlations are calculated: $A \cdot B^* = (S+N_1) \cdot (S^*+N_2^*) = S^2 + S \cdot N_2^* + N_1 \cdot S^* + N_1 \cdot N_2^*$. The following ***FSC*** curves are generated (function of radius *r*):

1) A $FSC_{AB}$ (**Fig S1c**) between the two noisy volumes ***A*** and ***B***,
2) The *signal-versus-noise* correlation $FSC_{SN1}$ between ***S*** and $N_1$ (**Fig S1i**),
3) The *signal-versus-noise* correlation $FSC_{SN2}$ between ***S*** and $N_2$ (**Fig S1j**), and,
4) The *noise-versus-noise* correlation: $FSC_{N1N2}$, between $N_1$ and $N_2$ (**Fig S1h**).

The $FSC_{SN1}$, $FSC_{SN2}$, and $FSC_{N1N2}$ curves are normalized only by the number of voxels in the shells and are directly comparable. The $FSC_{AB}$ (**Fig S1c**) has a typical ***FSC*** appearance indicating that realistic levels of noise have been added to the signal ***S*** to create the noisy volumes ***A*** and ***B***. The noise-versus-noise $FSC_{N1N2}$ (**Fig S1h**) oscillates around the zero mark over all spatial frequencies with its modulation increasing close to the origin, where Fourier shells contain fewer voxels. The correlation fluctuations between the signal ***S*** and each of the noise terms $N_1$ and $N_2$ are significantly *larger* than are the correlations between the noise terms $N_1$ and $N_2$ in the relevant frequency ranges. Whenever the signal ***S*** and the uncorrelated noise $N_i$ are of similar size, the predominant noise component consists of the cross-terms ($S \cdot N_2^* + N_1 \cdot S^*$) rather than of the smaller ($N_1 \cdot N_2^*$) correlations. Resolution criteria must compare the overall signal component to the influence of *all three* noise-related cross-terms. It is at this level that a fundamental error is often made in the literature. It is, namely assumed that ***because*** the signal and the noise are ***independent*** or ***uncorrelated***, their inner products $S \cdot N_2^*$ and $N_1 \cdot S^*$ are ***zero*** and may be dropped from the equations!

However, ***uncorrelated*** in this context means only that the inner product between these vectors is zero as an ***expectation value***, that is, as the average over an infinite number of repeats of the experiment (symbolically: $<S \cdot N> = 0$). Assuming that ***each*** of the individual $S \cdot N$ inner products is zero implies, in contrast, that ***each*** of these vector pairs are ***orthogonal***. The latter assumption – prolific in the EM literature – is untenable as is clearly illustrated in our model experiment. Thus, although the statement that the *s*ignal **S** and the noise **N** are *independent* or *uncorrelated* is by itself correct, the "standard interpretation" of its practical consequence in EM, namely, that each of the signal and the noise vector-pairs are *orthogonal*, is fundamentally flawed.

Note that various definitions of *orthogonal*, *uncorrelated*, or *independent* vectors exist, which confusion extends to most fields of science. The same confusion exists in the use of these concepts in our daily language. In mathematics, for example, when the inner product of two vector is *zero*, those vectors are called *orthogonal*. In statistics, on the other hand, the same *orthogonal* nomenclature is often used for when the *expectation value* of the inner product is *zero*. One must thus clearly define what exactly one means with any such statistical statement and one must stay within one self-consistent set of definitions.

Figure S1: Fourier Shell Correlations twixt Signal and Noise volumes.
   a) Section through test volume **A**, the sum of noise-free signal/volume **S** (d) and random noise volume **N₁** (f). Shown is section 151 (of 360).
   b) Corresponding section through test volume **B**, the sum of signal **S** (d) and a second random noise volume **N₂** (g).
   c) *FSC* between the two noisy volumes **A** and **B**.
   d) Noise-free signal/volume S (an arbitrary cryo-EM 3D reconstruction). Shown is section 151 (of 360).
   e) Copy of the same noise-free signal/volume **S** (d).
   f) Corresponding section through noise volume **N₁**.
   g) Corresponding section through noise volume **N₂**.
   h) *FSC* (not normalised) between noise volumes **N₁** and **N₂**.
   i) *FSC* (not normalised) between signal **S** and noise volume **N₁**.
   j) *FSC* (not normalised) between signal **S** and noise volume **N₂**.

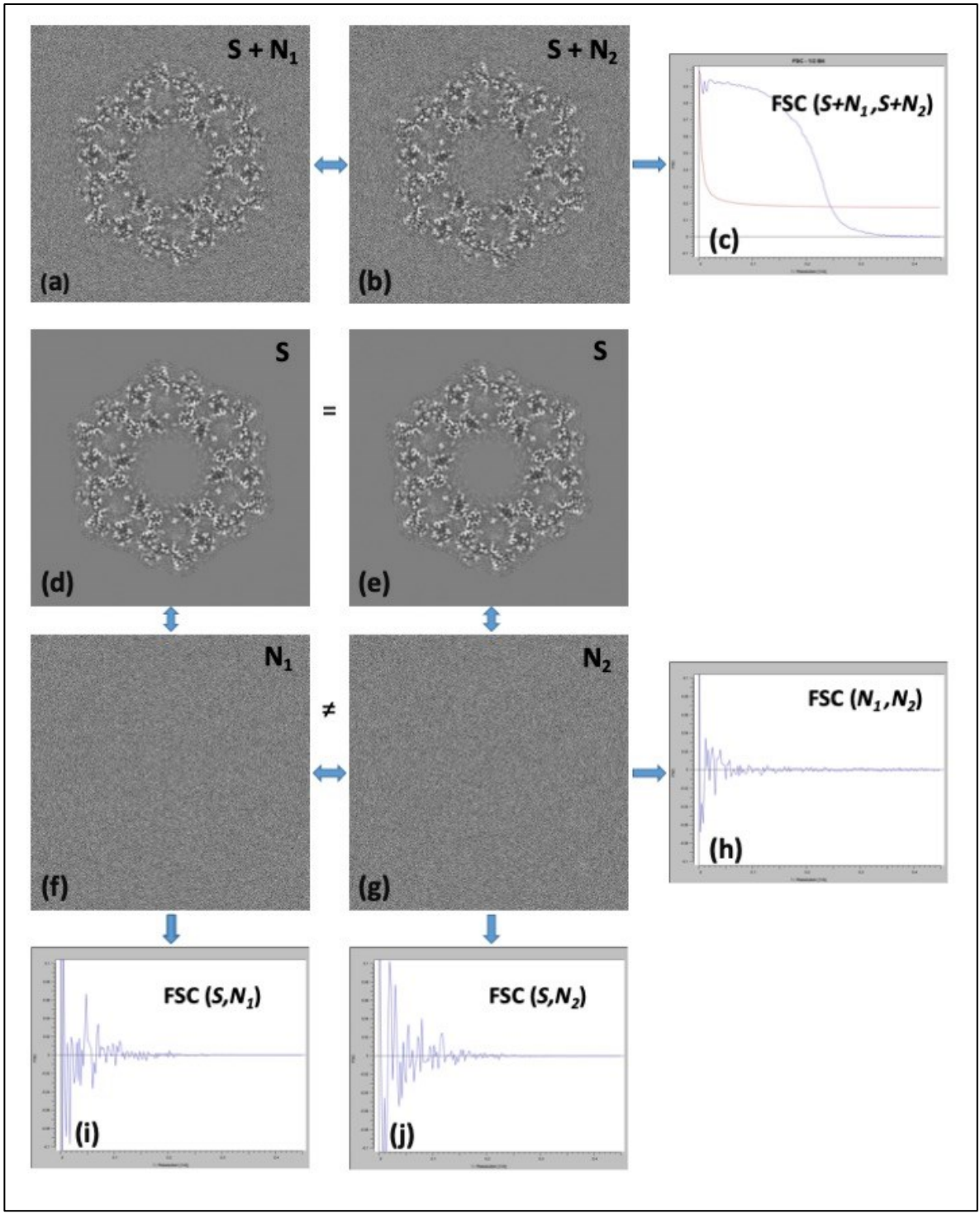